\documentclass[pdflatex,sn-mathphys-ay]{sn-jnl}
\usepackage{graphicx}
\usepackage{xurl}
\usepackage{multirow}
\usepackage{booktabs}
\usepackage{tabularx}
\usepackage{makecell}
\usepackage{rotating}
\usepackage{placeins}
\usepackage{amsmath,amssymb,amsfonts}
\usepackage{amsthm} 
\usepackage{mathrsfs}
\usepackage{manyfoot}
\usepackage{xcolor}
\theoremstyle{thmstyleone}

\theoremstyle{thmstyletwo}

\theoremstyle{thmstylethree}

\begin{document}

\title[Benchmarking Modular Optimization Strategies for Parameterized Quantum Circuits]{Benchmarking Modular Optimization Strategies for Parameterized Quantum Circuits}

\author*[1]{\fnm{Carla} \sur{Cotea}}\email{coteacotea@gmail.com}
\author[1, 2]{\fnm{Stefan} \sur{Balauca}}\email{stefan-razvan.balauca@info.uaic.ro}
\author[1]{\fnm{Andreea} \sur{Arusoaie}}\email{andreea.arusoaie@info.uaic.ro}

\affil[1]{\orgdiv{Department of Computer Science}, \orgname{Alexandru Ioan Cuza University}, \city{Iași}, \country{Romania}}
\affil[2]{\orgname{FreeYa Mind Campus}, \city{Iași}, \country{Romania}}

\abstract{Training parameterized quantum circuits requires balancing optimization progress with limited evaluation budgets and the statistical uncertainty of near-term quantum hardware. We present a modular benchmarking framework that separates quantum search-direction estimation from classical parameter-update rules, allowing their interactions and sensitivities to be examined independently. We evaluate a suite of gradient-based, stochastic, and derivative-free optimizers across combinatorial optimization using the Quantum Approximate Optimization Algorithm (QAOA), supervised quantum machine learning using Iris classification and a quantum convolutional neural network (QCNN) for binary MNIST classification, and quantum chemistry using the variational quantum eigensolver (VQE) for molecular hydrogen. By separating objective evaluations from sample-circuit costs under finite-shot simulation and physical execution on a 156-qubit processor, we compare optimizer behavior across workloads with different parameter counts and measurement requirements. Four selected simulator and hardware case studies report terminal and best observed objectives separately, alongside post-training classifier accuracies. The comparisons include selected MaxCut and hydrogen trajectories and illustrate workload-dependent objective changes and evaluation costs, while the simulator benchmark compares both peak and mean performance across three seeds. The hardware runs do not establish an optimizer ranking or isolate a causal effect of device noise.}


\maketitle

\section*{Acknowledgments}
This work was partially supported by RExQTCS: Romanian Excellence Quantum Technologies enhancing Cybersecurity, a grant of the Ministry of Education and Research, CCCDI-UEFISCDI, project number PN-IV-P6-6.1-CoEx-2024-0214, within PNCDI IV.

We gratefully acknowledge FreeYa Mind Campus for providing educational access to the IBM Quantum resources used to conduct the physical hardware experiments presented in this work. The views expressed are those of the authors and do not reflect the official policy or position of IBM or the IBM Quantum team.

\section{Introduction}
\label{sec:intro}

Near-term quantum processors provide growing computational capabilities but
remain constrained by imperfect operations, decoherence, and the depth of
circuits that can be executed reliably
\citep{Preskill2018NISQ,Bharti2022Noisy}. Variational quantum algorithms (VQAs)
address this operating regime through a hybrid loop in which a quantum
processing unit evaluates a parameterized quantum circuit (PQC) and a classical
processor updates its parameters \citep{Cerezo2021Variational}. This pattern
supports applications including VQE for ground state estimation
\citep{peruzzo2014variational,tilly2022variational}, QAOA for combinatorial
optimization \citep{farhi2014quantum}, and variational quantum-learning models
such as QCNNs \citep{sim2019expressibility,cong2019quantum}.

\begin{figure}[t]
    \centering
    \includegraphics[width=\textwidth]{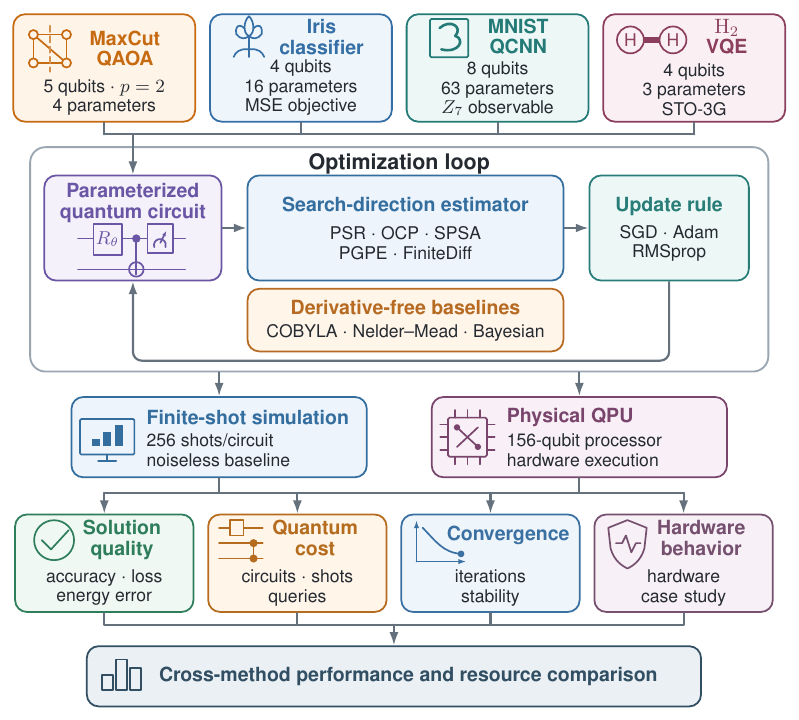}
    \caption{Overview of the benchmarking framework. Four parameterized-quantum-circuit workloads are evaluated using a modular optimization loop that separates search-direction estimation from classical parameter updating. Experiments are conducted under finite-shot simulation and selected physical-QPU execution, with comparisons based on solution quality, quantum-evaluation cost, convergence behavior, and hardware-case-study outcomes. Derivative-free methods retain their native proposal and update dynamics}
    \label{fig:benchmark_overview}
\end{figure}

Their practical performance, however, depends on more than the expressivity of
the circuit. Optimization must navigate non-convex landscapes in which gradients
can become exponentially suppressed \citep{McClean2018Barren,cerezo2021cost},
while finite-shot measurements make objective and gradient estimates stochastic.
Physical noise can further flatten the landscape as circuit depth or system size
increases \citep{Wang2021NoiseInduced}. Comparative studies have consequently
evaluated gradient-based, stochastic, and derivative-free optimizers under ideal,
finite-shot, and device-informed conditions
\citep{Lavrijsen2020Classical,PellowJarman2021Comparison,
BonetMonroig2023Performance,Singh2023Benchmarking}. Their findings consistently
show that optimizer performance depends on the workload, sampling budget, noise
model, and hyperparameter choices. Yet most comparisons treat each optimizer as
a complete package. They reveal which implementation performs well in a given
experiment, but provide less insight into why its behavior changes across
settings.

For gradient-based and stochastic methods, the outer loop contains two
conceptually distinct operations. Quantum evaluations are first combined into
an estimated search direction; a classical update rule then transforms that
direction using a learning rate, momentum, or adaptive moment estimates. A poor
trajectory can therefore arise from the bias or variance of the direction
estimator, from the way the classical update processes that direction, or from
their interaction. This distinction is particularly important under finite-shot execution because parameter-shift, coordinate-wise, simultaneous-perturbation,
and distributional estimators incur different sampling costs and respond
differently to measurement noise
\citep{Schuld2019Evaluating,Stokes2020Quantum,sweke2020stochastic}.

We study this distinction by independently pairing parameter-shift, finite-
difference, SPSA, PGPE, and diagonal observable-curvature-preconditioned
directions with SGD, Adam, and RMSprop. COBYLA, Nelder--Mead, and Bayesian
optimization are retained as derivative-free baselines with their native search
and update dynamics. The configurations are evaluated on four workloads: a
five-qubit MaxCut QAOA instance, a four-qubit Iris classifier, an eight-qubit
and 63 parameter QCNN for binary MNIST classification, and a four-qubit
$\mathrm{H}_2$ VQE problem. All workloads are first executed with noiseless
circuit evolution and 256-shot measurement sampling; selected configurations
are subsequently examined on the 156 qubit \texttt{ibm\_aachen} processor.

This study makes three main contributions. First, it operationalizes the
separation between search-direction estimation and classical parameter updating,
allowing differences between estimator and update rule combinations to be examined explicitly.
Second, it compares these combinations across optimization, classification, and
quantum-chemistry workloads rather than deriving conclusions from a single VQA
instance. Third, it distinguishes complete objective evaluations from logical
circuit executions, exposing costs introduced by parameter count, mini batches,
shifted evaluations, and observable measurement. The hardware experiments serve
as targeted case studies of selected simulation findings, not as a general test
of simulator-to-device transfer or quantum advantage.

The selected runs show objective improvement on both backends, with different terminal and best observed values. Iris COBYLA uses relatively few sample-circuit evaluations, whereas QCNN OCP-RMSprop requires shifted measurements for every parameter. The $H_2$ PGPE-SGD trajectory exhibits transient hardware improvement followed by deterioration. All MaxCut runs use the same graph generated with seed 42, independently of the optimizer seed. These observations describe the selected configurations and budgets; they do not provide a universal optimizer ranking or isolate a hardware noise mechanism.

Figure~\ref{fig:benchmark_overview} provides a compact overview of the benchmark design, from the four workload families through the modular optimization loop, execution environments, and reported evaluation dimensions.

The remainder of the paper is organized as follows.
Section~\ref{sec:background_related} introduces the VQA optimization loop and
positions the study against comparative and resource-aware work.
Section~\ref{sec:benchmark_design} defines the component-level comparison,
evaluation scope, and resource-accounting convention, while
Section~\ref{sec:experimental_methodology} describes the workloads, optimizer
implementations, and execution protocols. The experimental results are presented and interpreted
in Section~\ref{sec:results}. Study limitations are discussed
in Section~\ref{sec:discussion}. Section~\ref{sec:conclusion} concludes the paper
and outlines directions for future work.

\section{Background and Related Work}
\label{sec:background_related}

This section introduces the optimization concepts needed to interpret the
benchmark and places the study in the context of prior work. We first
distinguish search-direction estimation from classical parameter updating. We
then review comparative and resource-aware studies before identifying the
component-level gap addressed in this work.

\subsection{VQA Optimization Background}
\label{subsec:vqa_optimization_background}

A VQA minimizes an objective $J(\boldsymbol\theta)$ estimated by measuring a
parameterized quantum state. At iteration $k$, a direction estimator uses these
measurements to construct $\widehat{\mathbf d}_k$. Depending on the method, this
may be an estimate of $\nabla J(\boldsymbol\theta_k)$, a preconditioned gradient,
or the gradient of a smoothed objective. A classical update rule transforms this
direction into $\mathbf p_k$ and advances the parameters as
\begin{align}
    (\mathbf p_k,\mathbf s_{k+1})
        &=\operatorname{Update}(\widehat{\mathbf d}_k,\mathbf s_k),
        \label{eq:background_update_direction}\\
    \boldsymbol\theta_{k+1}
        &=\boldsymbol\theta_k-\eta_k\mathbf p_k,
        \label{eq:background_parameter_update}
\end{align}
where $\eta_k$ is the learning rate and $\mathbf s_k$ represents optimizer state,
such as momentum or second-moment estimates. SGD uses
$\mathbf p_k=\widehat{\mathbf d}_k$ directly. Adam and RMSprop instead combine
the current direction with information accumulated over earlier iterations.
The decomposition therefore separates the information obtained from quantum
evaluations from the classical dynamics of the parameter update.

Different estimators obtain this direction from different quantum evaluations.
Parameter-shift rules provide analytic derivatives for suitable gates
\citep{Schuld2019Evaluating}, whereas finite differences approximate them
numerically \citep{kiefer1952stochastic}. SPSA perturbs all parameters
simultaneously \citep{spall1992multivariate}, while PGPE estimates the gradient
of a smoothed objective through parameter-space exploration
\citep{sehnke2010parameter}. Derivative-free algorithms couple their choice of
query points to their own update dynamics and are therefore treated as complete
optimizer baselines.

Under finite-shot execution, these choices affect both precision and quantum
cost. A single objective evaluation may span a mini batch or several measurement
settings, and one direction estimate may require several such evaluations.
Section~\ref{sec:benchmark_design} defines the distinction between objective
evaluations, logical circuit executions, and shots, while
Section~\ref{sec:experimental_methodology} gives the method-specific
details.
\subsection{Comparative and Resource-Aware Studies}
\label{subsec:related_optimizer_benchmarks}
\label{subsec:related_quantum_aware_optimization}

Existing literature provides little evidence for a universal optimizer ranking. Previous benchmarks spanning combinatorial optimization, VQE, and linear solvers have tested gradient-based, stochastic, and derivative-free methods under ideal, finite-shot, and hardware-informed conditions \citep{Nannicini2019Performance,Lavrijsen2020Classical, PellowJarman2021Comparison,Singh2023Benchmarking}. Recent evaluations also include metaheuristics, comparing population-based methods such as the Covariance Matrix Adaptation Evolution Strategy (CMA-ES), Differential Evolution, and Particle Swarm Optimization with standard baselines on noisy VQA landscapes \citep{BonetMonroig2023Performance}. PGPE provides a direction estimator based on exploration in parameter space \citep{sehnke2010parameter}. Across these studies, optimizer performance is consistently reported as sensitive to the objective, circuit structure, noise profile, and parameter count.

Experimental protocols heavily influence these comparisons. As \citet{BonetMonroig2023Performance} demonstrated, hyperparameter tuning and sampling schedules can substantially alter the apparent ranking of optimizers. Similar final objective values may also require substantially different measurement budgets. This highlights the need for benchmarks that measure both solution quality and the underlying quantum resource costs.

Another line of work focuses on shot allocation inside the optimization loop. The precision of parameter-shift gradients can be directly traded against measurement counts \citep{Mari2021Estimating}. Adaptive methods like iCANS and Rosalin extend this concept by dynamically distributing shots across specific coordinates, iterations, or Hamiltonian terms \citep{Kubler2020Adaptive,Arrasmith2020Operator}.

Geometry-aware methods aim to improve the search direction directly. For instance, the quantum natural gradient relies on the Fubini-Study metric to capture the local geometry of the parameterized state \citep{Stokes2020Quantum}. Since computing the exact metric is expensive, later variants introduce stochastic approximations or reuse metric information across steps \citep{Gacon2021SPSA}. While geometry-aware, evolutionary, and metaheuristic approaches all show how gradient estimation, resource allocation, and update rules interact, the literature typically evaluates them as monolithic algorithms. Treating these solvers as opaque, coupled procedures makes it difficult to trace performance gains to a specific component of the optimization loop.

The gap addressed here is the comparison of independently paired direction
estimators and classical update rules across workloads, with objective queries
and logical circuit costs reported separately. Section~\ref{sec:benchmark_design}
defines the configurations and scope of this comparison.

\section{Benchmark Design}
\label{sec:benchmark_design}

\subsection{Optimization Components and Evaluation Scope}
\label{subsec:benchmark_modular_loop}
\label{subsec:benchmark_scope}
\label{subsec:related_study_position}

Following the decomposition in Eqs.~\eqref{eq:background_update_direction}
and~\eqref{eq:background_parameter_update}, we pair PSR, finite differences,
SPSA, PGPE, and observable-curvature preconditioning (OCP) with SGD, Adam,
and RMSprop. The update rules follow \citet{Robbins1951Stochastic},
\citet{Kingma2015Adam}, and \citet{Tieleman2012RMSProp}, respectively.
COBYLA, Nelder--Mead, and Bayesian optimization retain their native proposal
and update dynamics; imposing an external update rule would alter these methods.
Table~\ref{tab:optimization_components} summarizes the compared components.

The four workloads have different parameter counts: four for MaxCut QAOA,
16 for Iris, 63 for the binary-MNIST QCNN, and three for $H_2$ VQE.
We examine estimator--update interactions and achieved quality relative to
evaluation cost under 256 shots per setting. Parameter count, architecture,
and task change together, so cross-task trends are associations rather than
controlled effects of dimensionality. Selected hardware executions provide
single-trajectory case studies rather than an end-to-end solver comparison
or a test of quantum advantage.

Simulator configurations use seeds 42, 123, and 1234. The appendix reports the
100-step modular checkpoint or earlier termination. Selected simulator and
hardware comparisons use 25 steps for MaxCut and $H_2$, ten QCNN updates, and
50 Iris objective queries. Derivative-free methods retain their native query
and stopping semantics, detailed in Section~\ref{sec:experimental_methodology}.

\begin{table}[t]
    \centering
    \caption{Optimization strategies included in the benchmark. Modular
    direction estimators are paired with each external update rule, whereas
    derivative-free baselines retain their native proposal and update dynamics.}
    \label{tab:optimization_components}
    \footnotesize
    \begin{tabular}{@{}lllc@{}}
        \toprule
        \makecell[l]{\textbf{Method and}\\\textbf{reference}} &
        \textbf{Role} &
        \makecell[l]{\textbf{Nominal objective}\\\textbf{evaluations per step}} &
        \makecell[l]{\textbf{Parameter-update}\\\textbf{mechanism}} \\
        \midrule
        \makecell[l]{PSR\\[-1pt]\citep{Schuld2019Evaluating}} &
        Coordinate gradient & \makecell[l]{$1+2M$ or\\gate-dependent} & \makecell{SGD, Adam,\\RMSprop} \\
        \addlinespace[2pt]
        \makecell[l]{Observable-curvature\\preconditioning} &
        Scaled gradient & \makecell[l]{$1+4M$ or\\gate-dependent} & \makecell{SGD, Adam,\\RMSprop} \\
        \addlinespace[2pt]
        \makecell[l]{Finite differences\\[-1pt]\citep{kiefer1952stochastic}} &
        Coordinate gradient & $1+2M$ & \makecell{SGD, Adam,\\RMSprop} \\
        \addlinespace[2pt]
        \makecell[l]{SPSA\\[-1pt]\citep{spall1992multivariate}} &
        Simultaneous direction & $3$ & \makecell{SGD, Adam,\\RMSprop} \\
        \addlinespace[2pt]
        \makecell[l]{PGPE\\[-1pt]\citep{sehnke2010parameter}} &
        Smoothed direction & $1+2N$ & \makecell{SGD, Adam,\\RMSprop} \\
        \addlinespace[2pt]
        \makecell[l]{COBYLA\\[-1pt]\citep{powell1994direct}} &
        Complete optimizer & Native query sequence & Internal \\
        \addlinespace[2pt]
        \makecell[l]{Nelder--Mead\\[-1pt]\citep{nelder1965simplex}} &
        Complete optimizer & Native query sequence & Internal \\
        \addlinespace[2pt]
        \makecell[l]{Bayesian optimization\\[-1pt]\citep{frazier2018tutorial}} &
        Complete optimizer & One per acquisition & Internal \\
        \bottomrule
    \end{tabular}%
    \begin{flushleft}
        \footnotesize
        $M$ denotes the number of trainable parameters and $N$ the number of
        antithetic PGPE perturbation pairs. Counts include one unperturbed
        monitoring objective where used. For classification, logical circuit
        counts additionally scale with mini batch size $B$; PSR and curvature
        costs for energy workloads depend on the parameterized-gate structure.
        The observable-curvature row denotes the implemented heuristic, not an
        estimate of the Fubini--Study metric or QFIM.
    \end{flushleft}
\end{table}

\subsection{Resource Accounting}
\label{subsec:resource_accounting}

An objective evaluation is one complete evaluation of a loss or energy
at a specified parameter vector. A logical circuit evaluation is one
circuit execution for one data sample and one required observable setting. A
classification objective evaluated on a mini batch of size $B$ therefore
requires at least $B$ logical circuit evaluations. For Hamiltonians decomposed
into non-commuting Pauli groups, including VQE, one objective evaluation requires
one circuit per measurement group. Total shots introduce a further multiplier
and may also include calibration and mitigation overhead on hardware.

Including the unperturbed monitoring objective, coordinate-wise FD uses
$1+2M$ objective evaluations per iteration, SPSA uses three, and antithetic PGPE
uses $1+2N$. Classification multiplies these counts by $B$. Its parameter-shift
and diagonal-curvature strategies use $B(1+2M)$ and $B(1+4M)$ logical circuit
evaluations, respectively. For energy objectives, parameter-shift costs depend
on the number of parameterized elementary gates, while curvature estimation also
requires mixed-gate shifts; identical configurations are reused where possible.

The resource tables distinguish objective queries from logical circuits.
One MaxCut query uses one logical circuit. Classifier query counts are
multiplied by the number of samples evaluated: 16 for both Iris and QCNN
training objectives, including derivative-free QCNN queries. Requested shots are the
logical-circuit count multiplied by 256. For VQE, the logs retain objective
counts but not the number of commuting measurement groups, so logical-circuit
and shot totals are reported as not recorded. Initial reference evaluations,
post-training evaluations, calibration, mitigation, and service overhead are
outside the tracked training budget.

The column ``Steps/Queries'' counts entries in the stored history: one monitored step for a modular method and one objective query for COBYLA,
Nelder--Mead, or Bayesian optimization. The SciPy wrappers append a history
entry inside each objective call, not in a solver callback. Consequently,
Nelder--Mead can record more queries than its configured solver-iteration
limit. The Bayesian implementation stores each raw objective value in
\texttt{history}, aligned with its evaluated parameters, and records the running
minimum separately in \texttt{best\_history}. Both include initialization queries.
Trajectory diagnostics use the raw history. These counts are not a common iteration unit across methods.

\section{Experimental Methodology}
\label{sec:experimental_methodology}

This section specifies the benchmark workloads, implemented optimization
methods, data preparation, and simulation and hardware protocols. The conceptual
comparison axes and resource-accounting convention are defined in
Section~\ref{sec:benchmark_design}.

\subsection{Benchmark Workloads}
\label{subsec:experimental_workloads}

\noindent \textbf{MaxCut QAOA.}
We use a five-qubit Quantum Approximate Optimization Algorithm (QAOA)
\citep{farhi2014quantum} ansatz with $p=2$ and four trainable angles on the
unweighted graph
\[
 V=\{0,1,2,3,4\},\qquad
 E=\{(0,1),(0,3),(0,4),(1,3),(1,4),(2,3)\}.
\]
The reference state and layer structure are
\begin{align}
 |\phi_0\rangle&=|+\rangle^{\otimes5}, \label{eq:maxcut_ref}\\
 U(\boldsymbol\theta)&=U_M(\beta_2)U_C(\gamma_2)
 U_M(\beta_1)U_C(\gamma_1), \label{eq:qaoa_ansatz}
\end{align}
with transverse-field mixer
$U_M(\beta)=\exp(-i\beta\sum_jX_j)$. The minimized objective is
\begin{equation}
 \hat H_{\mathrm{MaxCut}}=-\frac12\sum_{(j,k)\in E}(I-Z_jZ_k).
 \label{eq:maxcut_ham}
\end{equation}
Exhaustive search gives a maximum cut of five and hence the target energy
$-5.0$. The implementation defines
$U_C(\gamma)=\exp(-i\gamma\hat H_{\mathrm{MaxCut}})$ and minimizes
$\langle\hat H_{\mathrm{MaxCut}}\rangle$ consistently across estimators.

\noindent \textbf{Iris classification.}
For low-dimensional supervised learning, we use a hardware-efficient ansatz
(HEA) trained on the first two Iris classes. A min--max scaler fitted only
on the training partition maps its feature ranges to $[0,\pi]$ and applies
the same transformation to the test partition. The four features are encoded
on four qubits as
\begin{equation}
 |\psi(\mathbf x)\rangle=\bigotimes_{i=0}^3R_X(x_i)|0\rangle.
 \label{eq:iris_encoding}
\end{equation}
Two HEA layers apply independent $R_Y$ and $R_Z$ rotations to every qubit,
followed by the directed CNOT ring $0\to1$, $1\to2$, $2\to3$, $3\to0$.
The circuit has 16 trainable parameters and output observable $Z_0$.

For Iris and MNIST, the output expectation is mapped to a class-1 score,
\begin{equation}
 \widetilde y_{b,\boldsymbol\theta}=
 \frac{1+\langle Z_{\mathrm{out}}\rangle_{b,\boldsymbol\theta}}{2},
 \label{eq:probability_mapping}
\end{equation}
and class 1 is predicted when $\widetilde y\geq0.5$. Training minimizes
\begin{equation}
 \mathcal L_{\mathrm{MSE}}(\boldsymbol\theta)=
 \frac1B\sum_{b=1}^{B}
 (\widetilde y_{b,\boldsymbol\theta}-y_b)^2,
 \qquad y_b\in\{0,1\}.
 \label{eq:mse_loss}
\end{equation}

\noindent \textbf{MNIST classification.}
The higher-dimensional supervised-learning workload uses an eight-qubit QCNN.
MNIST digits 0 and 1 are reduced to eight features using resizing, Otsu
binarization, and spatial averaging. Phase encoding prepares
\begin{equation}
 |\psi(\mathbf x)\rangle=
 \left[\bigotimes_{j=0}^7P(2x_j)\right]
 H^{\otimes8}|0\rangle^{\otimes8}.
 \label{eq:mnist_ref}
\end{equation}
The QCNN architecture, based on \citet{cong2019quantum}, has three
convolution--pooling stages with active-wire counts $8\to4\to2\to1$, 63
independent trainable parameters, and output observable $Z_7$. Appendix
\ref{secE1} specifies the gate templates and wire schedule. The output mapping
and MSE objective are given by Eqs.~\eqref{eq:probability_mapping} and
\eqref{eq:mse_loss}.

\noindent \textbf{Molecular hydrogen VQE.}
The four-qubit $H_2$ workload uses VQE \citep{peruzzo2014variational} with the
STO-3G basis at bond length $0.7414$~\AA, Jordan--Wigner mapping, and a Qiskit
Nature UCCSD ansatz. The Hartree--Fock reference occupies logical qubits 0 and 2,
\begin{equation}
 |\phi_{\mathrm{HF}}\rangle=|0101\rangle_{q_3q_2q_1q_0},
 \label{eq:vqe_ref}
\end{equation}
and the formal UCC operator is
\begin{equation}
 U(\boldsymbol\theta)=
 \exp[\hat T(\boldsymbol\theta)-\hat T^\dagger(\boldsymbol\theta)].
 \label{eq:uccsd_operator}
\end{equation}
Ordered excitation evolutions implement three independent amplitudes for
$0\to1$, $2\to3$, and $(0,2)\to(1,3)$, without an additional spin-singlet
constraint. The minimized total energy is
$\langle\hat H_{\mathrm{el}}\rangle+E_{\mathrm{nuc}}$, with reference
$E_0=-1.1372701747$ Ha. Appendix~\ref{secA1} lists the generated Pauli
coefficients and nuclear-repulsion offset.

\noindent \textbf{Initialization.}
\label{secJ1}
Initial parameters are sampled independently using workload-specific ranges:
\begin{equation}
 \theta_j\sim
 \begin{cases}
 \mathcal U(0,0.02\pi), & \text{MaxCut and VQE},\\
 \mathcal U(0,2\pi), & \text{Iris and MNIST QCNN}.
 \end{cases}
 \label{eq:small_init}
\end{equation}
MaxCut and VQE therefore use random small-angle initialization, whereas the
classifiers sample the full nominal rotation-angle domain.
These choices do not constitute a controlled comparison of initialization
strategies or establish the absence of barren plateaus. Reference state
preparation and data encoding remain part of each circuit.

\subsection{Direction Estimators and Optimizers}
\label{subsec:experimental_optimizers}

Throughout this subsection, $J(\boldsymbol\theta)$ denotes the scalar objective
being minimized. For the two classification workloads,
$J=\mathcal L_{\mathrm{MSE}}$; for MaxCut and VQE, $J$ is the corresponding
energy objective. We use $h(\boldsymbol\theta)=\langle O\rangle_{\boldsymbol\theta}$
when a formula applies specifically to the expectation value of an observable.
Hats distinguish finite-shot estimates from their exact-expectation
counterparts, and $\mathbf e_i$ denotes the $i$th Euclidean basis vector. We
write $\widehat{\mathbf g}$ for an estimated gradient and
$\widehat{\mathbf d}$ for the direction passed to the classical update rule;
these coincide for the unpreconditioned gradient estimators.

\noindent \textbf{Parameter-shift rule.}
For a parameter appearing in one gate
$U(\theta_i)=\exp(-i\theta_iG_i)$ with generator eigenvalues $\pm1/2$, the
parameter shift rule (PSR) of \citet{Schuld2019Evaluating} gives the exact expectation value
derivative
\begin{equation}
 \frac{\partial h(\boldsymbol\theta)}{\partial\theta_i}=
 \frac{h(\boldsymbol\theta+\frac\pi2\mathbf e_i)-
 h(\boldsymbol\theta-\frac\pi2\mathbf e_i)}{2}.
 \label{eq:psr_formula}
\end{equation}
Finite-shot execution replaces the two expectation values by sample estimates,
and therefore yields a stochastic derivative estimate. Each trainable Iris and
QCNN angle appears in one Pauli rotation. For QAOA and UCCSD, parameterized
operations are decomposed into elementary rotations with affine angles
$\phi_\ell(\boldsymbol\theta)$; the implementation shifts one gate at a time
and applies the chain rule,
\begin{equation}
 \frac{\partial h}{\partial\theta_j}
 =\sum_\ell a_{\ell j}\frac{\partial h}{\partial\phi_\ell},
 \qquad
 a_{\ell j}=\frac{\partial\phi_\ell}{\partial\theta_j}.
 \label{eq:gatewise_psr}
\end{equation}

For classification, let
$z_b(\boldsymbol\theta)=\langle Z_{\mathrm{out}}\rangle_b$ and
$p_b(\boldsymbol\theta)=[1+z_b(\boldsymbol\theta)]/2$. On one shared mini batch,
the exact-expectation derivatives satisfy
\begin{align}
 \frac{\partial z_b}{\partial\theta_j}
 &=\frac{z_b(\boldsymbol\theta+\frac\pi2\mathbf e_j)-
 z_b(\boldsymbol\theta-\frac\pi2\mathbf e_j)}{2},\\
 \frac{\partial\mathcal L_{\mathrm{MSE}}}{\partial\theta_j}
 &=\frac1B\sum_{b=1}^{B}
 (p_b(\boldsymbol\theta)-y_b)\frac{\partial z_b}{\partial\theta_j}.
 \label{eq:classifier_psr}
\end{align}
The implementation replaces each expectation by its finite-shot estimate and
uses the resulting $\widehat{\nabla J}$ as the search direction. Thus, the chain
rule is applied to shifted output expectations rather than to a shifted
nonlinear MSE loss.

For energy tasks, composite parameterized operations are
decomposed into supported elementary rotations. Each gate occurrence receives
an independent auxiliary angle, even when several occurrences depend on the
same trainable parameter. The constant Jacobian $a_{\ell j}$ is computed from
the original gate expressions; expressions with parameter dependent Jacobians
are rejected. At each iteration, the original parameters are bound to obtain
the auxiliary angles, and each required gate angle is shifted by $\pm\pi/2$.
Weighted expectation differences are then summed according to
Eq.~\eqref{eq:gatewise_psr}. This procedure accounts for parameters shared by
several QAOA or UCCSD gates. Shifting every occurrence simultaneously with the
single gate formula would not implement this chain rule.

For classifiers, one batch of 16 training samples is drawn without replacement
per direction estimate. The same samples are used for the unshifted circuit
and every shifted circuit. The unshifted expectations determine both the MSE
residuals and the monitored loss; the shifted expectations determine the
observable derivatives. All sample and shift combinations are submitted in a
batched estimator request. Thus, PSR requests $B(1+2M)$ sample expectations:
528 for Iris and 2032 for QCNN per iteration. Batching reduces submissions,
but does not reduce this logical measurement count. The unshifted result is
reused by the optimizer for monitoring, avoiding a second loss evaluation on a
different batch. The returned gradient is passed to the selected SGD, Adam, or
RMSprop update rule.

\noindent \textbf{Diagonal observable-curvature preconditioning.}
OCP rescales the PSR gradient using diagonal observable curvature. It does not
estimate the Fubini--Study metric or quantum Fisher information matrix (QFIM)
used in quantum natural-gradient methods \citep{Stokes2020Quantum}.

For energy workloads, its diagonal denominator is
\begin{equation}
 D_j=\left|\sum_{\ell,m}a_{\ell j}a_{mj}
 \frac{\partial^2 J}{\partial\phi_\ell\partial\phi_m}\right|
 +\epsilon_{\mathrm{reg}}.
 \label{eq:diagonal_qfim_main}
\end{equation}
Second derivatives are obtained by applying the gate-wise shift operator twice,
including mixed terms between gates controlled by the same ansatz parameter.
Let $\widehat g_j$ denote the corresponding finite-shot PSR estimate of
$\partial J/\partial\theta_j$. The returned direction has component
$\widehat d_j=\widehat g_j/D_j$. For classification, the denominator is instead
\begin{equation}
 D_j=\frac1B\sum_{b=1}^{B}\left|
 \frac{z_b(\boldsymbol\theta+\pi\mathbf e_j)+
 z_b(\boldsymbol\theta-\pi\mathbf e_j)-2z_b(\boldsymbol\theta)}{4}
 \right|+\epsilon_{\mathrm{reg}},
 \label{eq:classifier_metric}
\end{equation}
and the returned direction is again $\widehat d_j=\widehat g_j/D_j$, now with
$\widehat g_j$ obtained from Eq.~\eqref{eq:classifier_psr} after replacing the
expectations by their finite-shot estimates. Absolute values are taken before
averaging. This denominator is label-independent and is neither the MSE Hessian
nor its Gauss--Newton diagonal; Appendix~\ref{secF1} states the corresponding second derivative identities and
distinction.

The second derivative shifts follow the construction of
\citet{Mari2021Estimating}, who also study diagonal Newton optimization.
Our energy variant is related to that method but uses the positive denominator
$|\widehat{\partial_j^2J}|+\epsilon_{\mathrm{reg}}$ before applying the chosen
external update rule. Their diagonal Newton construction does not establish the
specific classifier denominator used here. For classification, averaging the
absolute observable curvatures before division is an implementation choice of
this benchmark.

Regularization is $0.001$ for Iris and $0.1$ for the other tasks. It bounds the
inverse diagonal scale by $1/\epsilon_{\mathrm{reg}}$, without guaranteeing
that a finite stochastic update decreases the objective. Adam and RMSprop
accumulate moments of the preconditioned direction, rather than of the raw PSR
gradient. Appendix~\ref{secF1} gives the explicit second derivative construction.

\noindent \textbf{Finite differences.}
The symmetric finite-difference (FD) estimate of the $i$th objective derivative
is
\begin{equation}
 \widehat g_i^{\mathrm{FD}}(\boldsymbol\theta)=
 \frac{\widehat J(\boldsymbol\theta+\delta_{\mathrm{FD}}\mathbf e_i)-
 \widehat J(\boldsymbol\theta-\delta_{\mathrm{FD}}\mathbf e_i)}
 {2\delta_{\mathrm{FD}}},
 \label{eq:finite_diff}
\end{equation}
with $\delta_{\mathrm{FD}}=0.1$ \citep{kiefer1952stochastic}. If each objective
estimate has variance $\sigma_J^2$, independent shifted estimates give
approximately
\begin{equation}
 \operatorname{Var}[\widehat g_i^{\mathrm{FD}}]
 \approx\frac{\sigma_J^2}{2\delta_{\mathrm{FD}}^2}.
 \label{eq:fd_variance_scaling}
\end{equation}
The denominator therefore amplifies the shot-noise contribution as
$\delta_{\mathrm{FD}}$ decreases.

\noindent \textbf{SPSA.}
SPSA perturbs all $M$ parameters along a Bernoulli vector
$\boldsymbol\Delta_k\in\{-1,+1\}^M$ \citep{spall1992multivariate}:
\begin{equation}
 \widehat g_{k,i}^{\mathrm{SPSA}}=
 \frac{\widehat J(\boldsymbol\theta_k+c_k\boldsymbol\Delta_k)-
 \widehat J(\boldsymbol\theta_k-c_k\boldsymbol\Delta_k)}
 {2c_k\Delta_{k,i}}.
 \label{eq:spsa_formula}
\end{equation}
It uses two perturbed objectives regardless of $M$, with an additional
unperturbed objective recorded for monitoring. SPSA supplies only the direction;
step scaling is delegated to SGD, Adam, or RMSprop.

\noindent \textbf{PGPE.}
PGPE is evaluated as a Gaussian-smoothed evolution-strategy direction estimator
\citep{sehnke2010parameter}. With antithetic perturbations
$\boldsymbol\xi_n\sim\mathcal N(\mathbf0,\sigma_{\mathrm{P}}^2I)$, it computes
\begin{equation}
 \widehat{\mathbf g}_{\sigma_{\mathrm{P}}}^{\mathrm{PGPE}}(\boldsymbol\theta)=
 \frac1N\sum_{n=1}^{N}
 \frac{[\widehat J(\boldsymbol\theta+\boldsymbol\xi_n)-
 \widehat J(\boldsymbol\theta-\boldsymbol\xi_n)]\boldsymbol\xi_n}
 {2\sigma_{\mathrm{P}}^2},
 \label{eq:pgpe_formula}
\end{equation}
where $\sigma_{\mathrm{P}}$ is the perturbation standard deviation. The
subscript indicates that this estimates the gradient of the Gaussian-smoothed
objective rather than, in general, $\nabla J$ itself. The
implementation uses $N=5$ antithetic pairs and one unperturbed monitoring
objective. PGPE is evaluated in simulation for every workload and with SGD in the $H_2$ VQE
hardware case. MaxCut hardware uses SPSA with SGD.

\noindent \textbf{Derivative-free baselines.}
COBYLA constructs a local linear approximation over a simplex and solves a
trust-region subproblem \citep{powell1994direct}. Nelder--Mead maintains a
simplex of $M+1$ vertices and proposes reflection, expansion, contraction, or
shrinkage steps according to ranked objective values
\citep{nelder1965simplex}. Bayesian optimization uses a Gaussian-process
surrogate with a Constant $\times$ Mat\'ern($\nu=2.5$) kernel and white-noise
component; candidates maximize expected improvement with $\xi=0.01$
\citep{frazier2018tutorial}. These methods retain their native update dynamics
and are not paired with SGD, Adam, or RMSprop.

\noindent \textbf{Classical update rules.}
SGD applies the estimated direction directly, RMSprop scales it using an
uncorrected second moment, and Adam combines first-moment momentum with
bias-corrected second-moment scaling. All use
\begin{equation}
 \eta_k=\frac{\eta_0}{1+\lambda k}.
 \label{eq:lr_decay}
\end{equation}
Appendix~\ref{secB1} gives the complete recurrences.

\subsection{Data Preparation}
\label{subsec:methods_data}
\label{secI1}

\noindent \textbf{Iris.}
The first two classes are retained, giving 100 samples. A stratified
$80\%$--$20\%$ split first produces 80 training and 20 test instances.
A min--max scaler is then fitted exclusively on the training features,
mapping their per-feature ranges to $[0,\pi]$. The fitted transformation
is applied to both partitions without refitting on the test data. Test
values outside the training feature ranges are not clipped and may map
outside $[0,\pi]$.

\noindent \textbf{MNIST.}
We take the first 100 images of each of digits 0 and 1 from the MNIST source
training set. Each image is normalized by 255, resized to $8\times8$, and
binarized using its Otsu threshold. Binary pixels are multiplied by $\pi/2$
and averaged across columns to give eight features in $[0,\pi/2]$. This
transformation fits no scaler to the combined dataset. An unstratified
$80\%$--$20\%$ split yields 160 training and 40 test instances; class
proportions depend on the split seed.

Both classification workloads use the mean squared error (MSE) objective defined in Eq.~\eqref{eq:mse_loss}. Modular methods use mini batches of size $B=16$; derivative-free sampling protocols are specified below. For consistency, the dataset partitions are generated once per global random seed and reused across optimizer configurations. The Iris train--test split is stratified by class, whereas the MNIST QCNN split is unstratified.

Before each classifier optimizer run, the NumPy and Python pseudo-random number generators (PRNGs) are reset to the global seed. This common initialization does not guarantee identical mini batch sequences across estimators. Because stochastic perturbation methods such as PGPE and SPSA share a PRNG stream with the batch sampler, the resulting batch sequences can diverge according to the number of random draws required by each estimator.

Batch management also varies across the evaluated algorithms. The generic objective function samples a new mini batch without replacement on every call. For perturbation-based evaluators, a single mini batch is drawn and shared across all parameter vectors evaluated during that call, whereas PSR and observable-curvature methods manage batching internally.

The derivative-free QCNN methods (COBYLA, Nelder--Mead, and Bayesian optimization) use mini batches of $B=16$ training samples per objective query. Each query draws a fresh batch without replacement from the 160-sample training set and evaluates its mean squared error in one estimator request. Each query therefore accounts for 16 sample-circuit evaluations, consistent with Table~\ref{tab:qcnn_resources}: COBYLA uses $595\times16=9{,}520$ circuits, Nelder--Mead uses $1{,}107\times16=17{,}712$, and Bayesian optimization uses $1{,}000\times16=16{,}000$. Iris derivative-free methods likewise draw a fresh mini batch of 16 samples for each objective query.

\subsection{Experimental Protocol}
\label{subsec:experimental_protocol}

\noindent \textbf{Bounds and boundary handling.}
\label{secH1}
The nominal domains are $[0,2\pi]$ for Iris, MaxCut, and QCNN rotations and
$[-\pi,\pi]$ for UCCSD amplitudes. Gradient-based methods, COBYLA, and
Nelder--Mead do not enforce these intervals. Bayesian optimization evaluates the supplied initial point without clipping
and subsequently restricts initialization and acquisition candidates to
$[-\pi,\pi]^M$ for every workload. This domain is an implementation choice.

\noindent \textbf{Hyperparameters.}
\label{subsec:experimental_hyperparameters}
Hyperparameters are fixed for the reported comparisons, with task-specific values stated below. The initial learning rate is $\eta_0=0.05$. We set $\lambda=0.05$ for Iris and $\lambda=0.01$ for MaxCut, QCNN, and VQE. Adam uses $(\beta_1,\beta_2)=(0.9, 0.999)$ and RMSprop uses a second-moment decay of $0.9$; both apply a $10^{-8}$ numerical stabilizer.

For the search-direction estimators, finite differences (FD) uses a step size of $0.1$. SPSA uses the decay schedule $c_k=0.1/(k+1)^{0.101}$. PGPE draws five antithetic perturbation pairs with a standard deviation of $\sigma_{\mathrm{P}}=0.1$. The observable-curvature regularizer is $0.001$ for Iris and $0.1$ for the remaining workloads. Each requested circuit uses 256 shots, and the classification mini batch size is 16.

Nelder-Mead initializes the simplex with coordinate increments of $0.2$. COBYLA retains SciPy defaults except for the evaluation limit. Bayesian optimization starts with five observations, including the initial point. It fits a Gaussian process with a constant-scaled Mat\'ern kernel ($\nu=2.5$) and a white-noise component. The expected-improvement acquisition function uses a $0.01$ offset. Gaussian-process hyperparameters are optimized with three restarts, and each acquisition step evaluates $\max(200, 50M)$ uniformly sampled candidates.

\noindent \textbf{Evaluation Limits and Checkpoints.}
The appendix benchmark configured modular estimators for at most 100 classical update steps. The tables report the best observed objective over the available trajectory and the associated classifier accuracies; resource tables separately report the terminal objective. These limits differ from the budgets of the selected simulator and hardware figures. If a run ends sooner, its available trajectory is used. For MaxCut, Iris, and VQE, COBYLA's evaluation limit and Nelder-Mead's iteration limit were both set to 100. Because QCNN iterations scale with dataset size, its COBYLA evaluation limit and Nelder-Mead iteration limit are multiplied by the number of batches: $100 \lfloor 160/16 \rfloor = 1000$. The QCNN resource table records 1000 Bayesian objective queries, including initialization. Derivative-free checkpoints record total objective queries rather than internal solver iterations.

Bayesian objective-query counts include initialization and acquisition
steps. Resource comparisons use the recorded query counts in the tables.

\noindent \textbf{QCNN losses and accuracies.}
QCNN training objectives are mini batch MSE values, including those recorded
for derivative-free methods. Best and terminal trajectory losses refer to the
recorded optimization objectives. Train and test accuracies are evaluated
separately at the selected parameters using the full training and test sets.
Post-training evaluation is excluded from the tracked training budget.

\noindent \textbf{Stopping, best-iterate selection, and trajectory diagnostics.}
\label{subsec:experimental_diagnostics}
The generic modular optimizer (FD, SPSA, PGPE, PSR, and OCP with
SGD, Adam, or RMSprop) defaults to a patience of ten consecutive comparisons.
For the selected simulator comparison functions, this patience is set above the
requested step budget, so all 25 MaxCut/$H_2$ steps and ten QCNN steps are recorded.
The following stopping rule describes the generic default, not those selected runs.
Let $\widehat J_k$ denote the monitored finite-shot objective at step $k$.
Execution ends when
\begin{equation}
 |\widehat J_k-\widehat J_{k-1}|<10^{-4}
\end{equation}
holds for ten consecutive comparisons, or when the configured iteration limit
is reached. A change at or above the tolerance resets the count. The first
objective provides the baseline, so ten comparisons require eleven recorded
objective values. The optimizer returns the parameters associated with the
lowest observed objective value.

COBYLA and Nelder--Mead retain their native SciPy termination criteria and
configured limits. Bayesian optimization terminates at its objective-evaluation
budget, including initialization; it has no patience-based criterion.

After each run, the tracker separately records the first adjacent loss change
below $10^{-4}$ as a retrospective trajectory diagnostic. This single-comparison
diagnostic does not establish stable convergence and is labelled
``first small-change iteration'' in the appendix resource tables.

\noindent \textbf{Hardware case studies.} Appendix~\ref{sec:hardware_logs} records the sessions, job counts, and
training-evaluation totals for the selected hardware runs.

Across the completed jobs of these four sessions, result metadata reports 256 shots per expectation setting, measurement mitigation and measurement twirling enabled, and gate twirling, dynamical decoupling, zero-noise extrapolation, and probabilistic error cancellation disabled. The reported shot field does not include a verified total for all service-side calibration and mitigation circuits. The calibration snapshot associated with the reported runs is summarized in Appendix~\ref{secC1}.

Each hardware configuration is represented by one trajectory. MaxCut's initial parameters match optimizer seed 123 on the fixed graph generated with seed 42. The other simulator comparisons use seed 42. Classical random seeds are not treated as Runtime metadata fields; the archived parameter histories and experiment configuration provide the relevant context.

\noindent \textbf{Finite-shot simulation.} The selected runs use \texttt{AerSimulator} without a device noise model, with 256 shots requested through Runtime Estimator V2 in local mode. NumPy, Python, Aer, and the level-3 transpiler use the selected task seed. The software versions pinned for these runs include Python 3.12, Qiskit 2.2.3, Qiskit Aer 0.17.2, Qiskit Nature 0.7.2, SciPy 1.17.1, and Qiskit IBM Runtime 0.43.1. Local execution has finite sampling but does not reproduce the hardware compilation, mitigation, or calibration environment.

\section{Results}\label{sec:results}

Performance is evaluated from two complementary perspectives: the peak result obtained by a configuration in any of the three seeds (123, 1234, and 42), and its mean performance across those seeds. A favorable run can produce an exceptional individual result without ensuring reliable performance across seeds. The comparisons below therefore distinguish the strongest observed outcome from the quality retained on average. The benchmark tables at the 100 step modular reporting point and native derivative-free limits in Appendix~\ref{secD1} provide the basis for these comparisons; the shorter simulator and hardware trajectories are reported separately.

For MaxCut, peak performance is the lowest reported best cost, and average performance is the mean of the three seed specific best costs. For VQE, best energy means the minimum sampled energy of the reported trajectory. Ranking uses the absolute difference between this energy and the FCI reference. Absolute errors are calculated before averaging; the trajectory is not searched for the energy closest to FCI. For classification, test accuracy is the primary metric and reported best loss is secondary. These averages describe three recorded runs, rather than a statistical guarantee of convergence. Classification seeds also change the data partition and stochastic training sequence, so their differences cannot be attributed exclusively to parameter initialization.

\begin{figure*}[t]
    \centering
    \includegraphics[width=\textwidth]{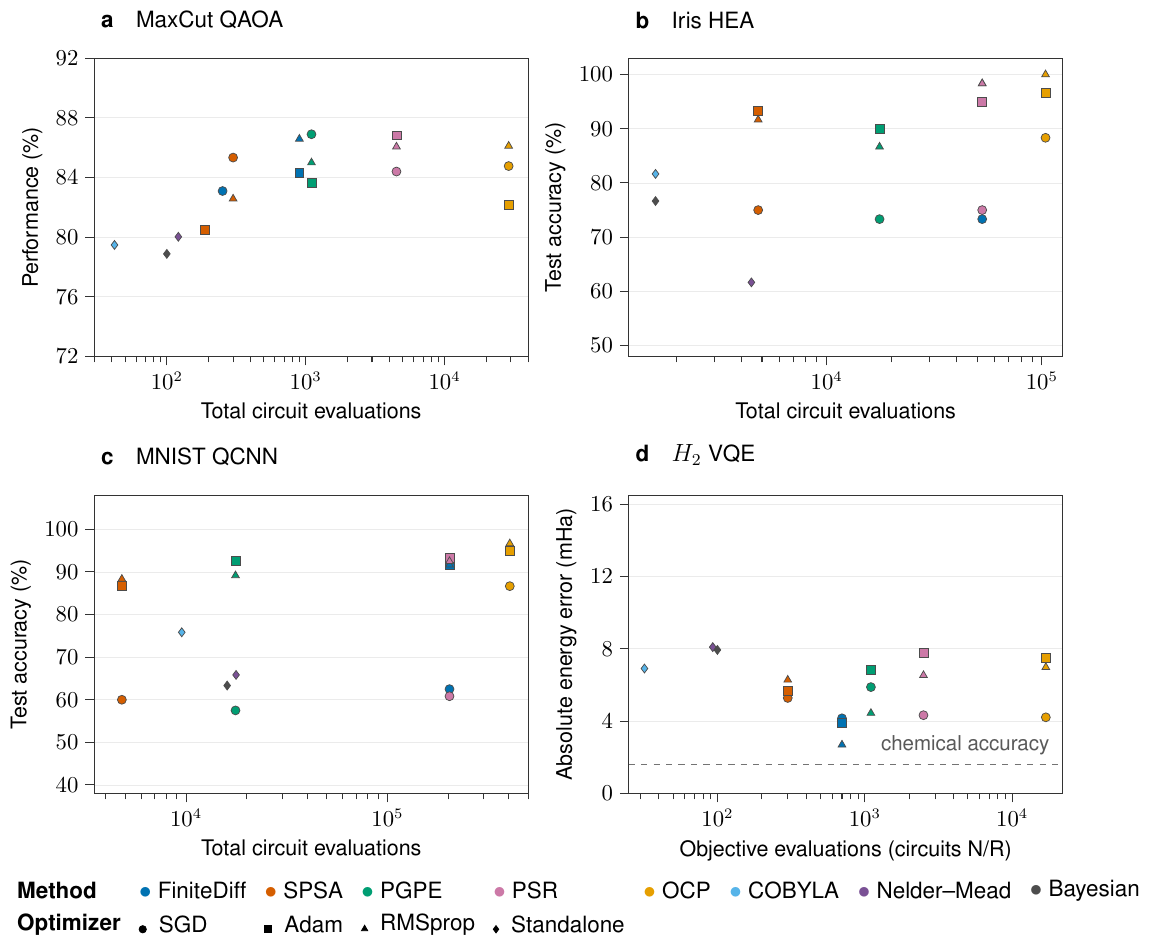}
    \caption{Mean performance versus evaluation cost across the four simulated workloads and three random seeds (42, 123, and 1234). Each marker denotes one optimizer configuration; marker color identifies the search-direction method and marker shape identifies the classical update rule or derivative-free baseline. MaxCut performance is expressed relative to the exact target cost, Iris and MNIST QCNN performance as test accuracy, and $H_2$ VQE performance as absolute error from the FCI reference energy. Logical-circuit evaluations are shown for MaxCut and the classification workloads. For VQE, only objective-evaluation counts are reported because the number of commuting measurement groups was not retained. Derivative-free methods follow their native query and stopping semantics, so horizontal positions should not be interpreted as equal iteration budgets across methods}
    \label{fig:performance_cost_summary}
\end{figure*}

Figure~\ref{fig:performance_cost_summary} summarizes the relationship between achieved performance and quantum-evaluation cost across the four simulated workloads. Each marker represents the mean outcome across seeds 42, 123, and 1234 for one optimizer configuration at the 100-step modular reporting point, or at the native query limit for derivative-free methods. The figure complements the task-specific tables by showing that similar or improved objective quality can require substantially different evaluation budgets, and that this trade-off depends strongly on both the workload and the optimization strategy. Because the workloads differ in circuit architecture, parameter count, batching, and observable structure, the panels should be interpreted independently rather than as a direct cross-task ranking.
\begin{table*}[t]
\centering
\caption{Recorded objectives for the selected simulator and hardware runs. Each cell reports terminal / best observed value; lower is better. Energy is in Ha for $H_2$, the MaxCut objective is the expected negative cut, and classifier objectives are mini batch MSE. Counts refer to optimizer steps except for COBYLA, where they count objective queries. No across-run uncertainty is estimated}
\label{tab:selected_run_results}
\footnotesize
\setlength{\tabcolsep}{4pt}
\begin{tabular}{@{}llrcc@{}}
\toprule
\textbf{Task} & \textbf{Optimizer} & \textbf{Count} & \textbf{Simulator: terminal / best} & \textbf{Hardware: terminal / best} \\
\midrule
MaxCut & SPSA-SGD & 25 & $-4.023438 / -4.027344$ & $-3.908474 / -3.984726$ \\
Iris & COBYLA & 50 & $0.059243 / 0.046108$ & $0.136762 / 0.091819$ \\
MNIST QCNN & OCP-RMSprop & 10 & $0.171905 / 0.165340$ & $0.181417 / 0.181417$ \\
$H_2$ VQE & PGPE-SGD & 25 & $-1.129593 / -1.138991$ & $-0.980468 / -1.063881$ \\
\bottomrule
\end{tabular}
\end{table*}

Table~\ref{tab:selected_run_results} distinguishes the terminal objective from the minimum observed along each selected trajectory. Classification accuracies are separate post-training evaluations of selected parameters. These single hardware trajectories are not across seeds averages and are not expected to reproduce the 100 step benchmark peaks.

The hardware cases were selected in light of the best individual outcomes across seeds, rather than the average-performance ranking: SPSA-SGD shares the MaxCut peak, PGPE-SGD gives the closest VQE estimate, and OCP-RMSprop shares the QCNN accuracy peak. The Iris case uses COBYLA, which shares the $100\%$ seed-42 test-accuracy peak but is not the average-performance leader. The selected methods are therefore not unique winners under a common criterion. Their shorter matched runs use 25, 50, 10, and 25 recorded steps or queries for MaxCut, Iris, QCNN, and VQE, respectively. Optimizer seeds are 123 for MaxCut and 42 for the other tasks, with the MaxCut graph fixed independently at seed 42. All runs request 256 shots per expectation setting.


\subsection{Combinatorial Optimization: MaxCut QAOA}
\label{subsec:results_hardware_maxcut}

At the 100 step checkpoint, SPSA-SGD and PSR-Adam share the lowest observed cost, $-4.378906$, both at seed 123 (Table~\ref{tab:maxcut_100_seed123}). This corresponds to a distance of $0.621094$ from the exact target $-5$ and a relative error of $12.42\%$. Their across seeds behavior differs, however. SPSA-SGD reaches only $-4.062500$ at seed 1234, whereas PSR-Adam remains between $-4.378906$ and $-4.300781$ across the three seeds.

As shown in Table~\ref{tab:maxcut_average}, PGPE-SGD provides the strongest average result, $-4.345052$, followed by PSR-Adam ($-4.341146$) and FiniteDiff-RMSprop ($-4.329427$). PGPE-SGD attains $-4.359375$, $-4.316406$, and $-4.359375$ at seeds 123, 1234, and 42, respectively, retaining near peak quality across initializations. In contrast, SPSA-SGD ranks sixth on average ($-4.266927$), despite sharing the single run optimum. PGPE-SGD therefore achieves the lowest mean cost among the evaluated configurations; this does not imply the smallest dispersion among all methods. Its mean advantage over PSR-Adam is only $0.003906$, and no significance claim is inferred from three seeds. All comparisons use the same graph generated with seed 42; the optimizer seed does not redefine the instance.

The selected hardware run is examined below.
\begin{figure}[t]
\centering
\includegraphics[width=84mm]{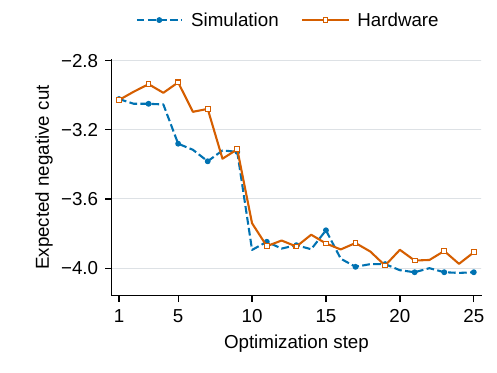}
\caption{MaxCut QAOA with SPSA-SGD over 25 monitored optimization steps on the same six-edge graph. Dashed lines with circles show simulation with optimizer seed 123 and graph seed 42; solid lines with squares show the hardware run on \texttt{ibm\_aachen}. The objective is expected negative cut, with exact target $-5$. The terminal simulator and hardware costs are $-4.023438$ and $-3.908474$; their best observed values are $-4.027344$ at step 24 and $-3.984726$ at step 19. The initial reference job and shifted gradient probes are excluded from the displayed history}
\label{fig:maxcut_spsa_hardware}
\end{figure}

Over 25 SPSA-SGD steps, the simulator objective changes from $-3.023438$ to $-4.023438$, with minimum $-4.027344$ at step 24. The hardware objective changes from $-3.028729$ to $-3.908474$, with minimum $-3.984726$ at step 19 (Fig.~\ref{fig:maxcut_spsa_hardware}). The terminal expected-cut ratios are therefore $80.47\%$ in simulation and $78.17\%$ on hardware, relative to the exact cut of five. The simulator has the lower terminal cost by approximately $0.114964$ on this matched instance. This is a difference between two selected trajectories, not a statistically estimated effect of device noise.

The batch contains the monitored center and two simultaneous-perturbation probes, yielding 75 objective evaluations during training. The hardware session contains 25 training jobs plus an initial reference job. The 24 observed parameter transitions agree with the implemented SPSA-SGD updates. Subsequent loss increases are compatible with the stochastic, non-monotone update rule; the hardware minimum at step 19 is not its terminal value.

\subsection{Quantum Chemistry: Molecular Hydrogen VQE}
\label{subsec:results_hardware_vqe}

The strongest individual agreement with the FCI reference is obtained by PGPE-SGD at seed 42: $-1.13728088$~Ha, corresponding to an absolute error of $0.0107$~mHa (Table~\ref{tab:vqe_100}). This is well within the nominal $1.6$~mHa chemical accuracy threshold. However, the absolute errors at seeds 123 and 1234 are $13.7041$ and $3.9440$~mHa, respectively. Its mean absolute error consequently increases to $5.8863$~mHa, placing it ninth in the across seeds comparison.

FiniteDiff-RMSprop achieves the lowest mean absolute error, $2.6949$~mHa, followed by FiniteDiff-Adam at $3.9324$~mHa (Table~\ref{tab:vqe_average}). The FiniteDiff-RMSprop errors are $7.1173$, $0.1723$, and $0.7952$~mHa across seeds 123, 1234, and 42. Thus, the method achieves the lowest mean absolute error among the evaluated configurations, although its mean error remains above the chemical accuracy threshold. Mean absolute error does not measure energy variance, and the aggregate ranking alone cannot establish minimum variance. Moreover, the reported best energies are finite-shot estimates: a value below the FCI reference is not evidence of a physically lower ground state energy. The exceptional PGPE-SGD peak should therefore be interpreted as observed numerical agreement, rather than independently certified chemical accuracy.

The selected hardware run is examined below.

\begin{figure}[t]
\centering
\includegraphics[width=84mm]{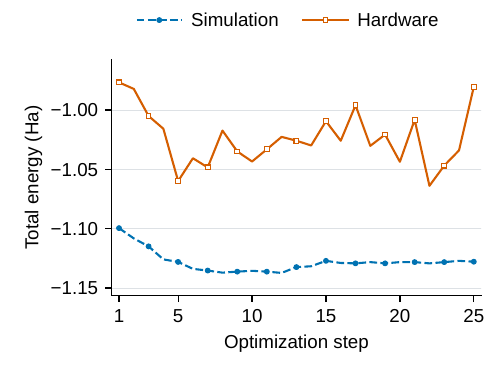}
\caption{$H_2$ VQE with PGPE-SGD: total energy over 25 monitored steps in seed-42 simulation (dashed lines with circles) and on \texttt{ibm\_aachen} (solid lines with squares). Both histories include nuclear repulsion and use 256 requested shots per expectation setting. Hardware points are the unshifted centers of the 11-evaluation PGPE batches executed on hardware. The terminal energies are $-1.129593$~Ha in simulation and $-0.980468$~Ha on hardware; the minimum measured energies are $-1.138991$ and $-1.063881$~Ha. The exact reference is $-1.137270174661$~Ha. Curves are unsmoothed single trajectories, not estimates of run-to-run uncertainty}
\label{fig:h2_pgpe_hardware}
\end{figure}

Figure~\ref{fig:h2_pgpe_hardware} compares the 25 step PGPE-SGD simulator run with the hardware session \nolinkurl{2e502439-33b1-460e-a070-f2122803e9fd}. The observable used in the hardware run was checked against the local four-qubit STO-3G electronic Hamiltonian under qubit permutation. Nuclear repulsion, $0.713753993688$~Ha, is added to both histories, giving a common total-energy convention and FCI reference $E_{\mathrm{FCI}}=-1.137270174661$~Ha. Each training batch comprises the unshifted center and five antithetic perturbation pairs. The plotted hardware values retain only the center; the initial reference and perturbation probes are excluded. All 24 recorded parameter transitions agree with PGPE-SGD using $\sigma=0.1$, learning rate $0.05$, and decay $0.01$.

The simulator starts at $-1.099531$~Ha and has terminal energy $-1.129593$~Ha, an absolute error of $7.6773$~mHa. Its minimum measured energy is $-1.138991$~Ha at step 8, with absolute deviation $1.7212$~mHa. This finite-shot observation lies below the FCI reference; it is not a physical improvement over the exact ground state. Selecting a minimum from noisy observations can favor downward fluctuations, and the minimum measured energy need not be the observation closest to the reference.

Hardware starts at $-0.976711$~Ha, reaches its lowest recorded energy of $-1.063881$~Ha at step 22, and ends at $-0.980468$~Ha. The corresponding best observed and terminal absolute errors are $73.3896$ and $156.8023$~mHa. The late deterioration means that the terminal energy alone hides the improvement achieved earlier. Neither the terminal nor minimum-energy estimates establish the nominal $1.6$~mHa chemical accuracy target. The larger hardware deviations are descriptive evidence for this run; attributing them to a particular gate, readout, or calibration effect would require additional controlled measurements.

\subsection{Supervised Quantum Machine Learning: Iris Classification}
\label{subsec:results_low_dim}

OCP-RMSprop achieves $100\%$ test accuracy at each of the three seeds, including seed 42, and consequently the highest mean test accuracy of $100\%$ (Tables~\ref{tab:iris_100_top} and~\ref{tab:iris_average}). Several configurations share a perfect single-seed score, but none of the others maintains it across all three runs. PSR-RMSprop and FiniteDiff-RMSprop follow at $98.33\%$ mean test accuracy, while OCP-Adam reaches $96.67\%$. The advantage of OCP-RMSprop is therefore expressed most clearly by its consistency across seeds, rather than by an exclusive single run maximum.

The loss ranking differs from the accuracy ranking. PSR-RMSprop has a slightly lower mean best loss ($0.030594$) than OCP-RMSprop ($0.030933$), despite its lower average test accuracy. At seed 42, PGPE-RMSprop records a still lower loss of $0.009272$ but only $90\%$ test accuracy. These results show why the lowest observed training mini batch loss is insufficient to identify the strongest classifier. The perfect OCP-RMSprop scores describe the evaluated small test partitions and do not establish perfect population-level generalization.

The selected hardware run is examined below.

\begin{figure}[t]
\centering
\includegraphics[width=84mm]{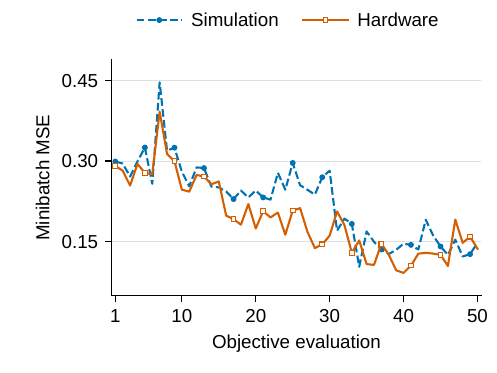}
\caption{Iris classification with COBYLA: simulator and hardware mini batch MSE over 50 objective queries. Dashed lines with circles denote the seed-42 simulator run; solid lines with squares denote the archived \texttt{ibm\_aachen} run. Each query uses 16 training samples and 256 requested shots per expectation setting. All recorded losses are plotted without smoothing; marker frequency is reduced for readability. The minima occur at simulator query 49 and hardware query 40. The post-training test accuracies of the selected parameters are $100\%$ and $90\%$, respectively, and are distinct from the plotted losses}
\label{fig:iris_hardware}
\end{figure}

For the four-qubit, 16-parameter Iris classifier, COBYLA records 50 objective queries, each using a newly drawn mini batch of 16 training samples. The simulator loss decreases from $0.291872$ to $0.059243$, with a minimum of $0.046108$ at query 49. The hardware loss decreases from $0.290979$ to $0.136762$, reaching its minimum of $0.091819$ at query 40 (Fig.~\ref{fig:iris_hardware}). 

Post-training evaluation of the selected simulator parameters gives $100\%$ training and test accuracy. The archived hardware evaluation gives $96.25\%$ training accuracy and $90\%$ test accuracy, corresponding to 77/80 and 18/20 correct classifications. The test-accuracy difference is therefore two samples. These observations support successful learning in both trajectories but do not establish a general hardware accuracy penalty. The mini batch composition changes between objective queries, so the loss curves do not measure full-training-set MSE at each step.

\subsection{Quantum Image Classification: MNIST QCNN}
\label{subsec:results_qcnn}

OCP-RMSprop also leads the eight-qubit QCNN benchmark by average test accuracy (Table~\ref{tab:qcnn_average}). Its seed specific accuracies are $97.5\%$, $92.5\%$, and $100\%$, giving a mean of $96.67\%$. At seed 42, the $100\%$ peak is shared with OCP-Adam (Table~\ref{tab:qcnn_100_top}); OCP-Adam ranks second on average at $95.00\%$, followed by PSR-Adam at $93.33\%$. PSR-RMSprop remains competitive at $92.50\%$, but its strongest result is $97.50\%$ at seed 123.

PSR-RMSprop achieves the lowest mean best recorded mini batch loss ($0.066232$), whereas OCP-RMSprop and OCP-Adam yield $0.067909$ and $0.067914$, respectively. The lower loss of PSR-RMSprop does not translate into the highest test accuracy. Across both classification tasks, OCP-RMSprop is therefore the leading configuration under the primary metric, while PSR-RMSprop and OCP-Adam offer competitive performance under complementary metrics. These are metric-specific empirical rankings, rather than dominance on every outcome.

The selected hardware run is examined below.

\begin{figure}[t]
\centering
\includegraphics[width=84mm]{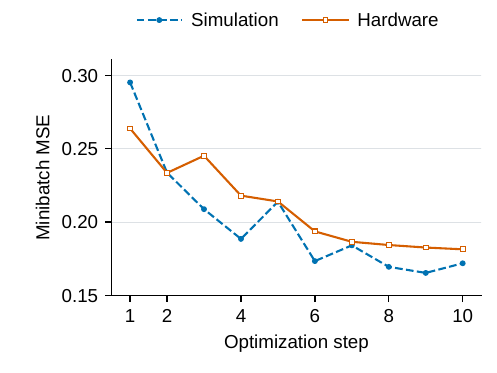}
\caption{Binary-MNIST QCNN optimization with OCP-RMSprop over ten mini batch updates. Dashed lines with circles show the seed-42 simulator trajectory and solid lines with squares show hardware observations on \texttt{ibm\_aachen}. Each monitored loss is MSE on a 16-sample training batch; 256 shots are requested per expectation setting. The simulator and hardware terminal losses are $0.171905$ and $0.181417$. Curves show individual observations without smoothing or across-run averaging; post-training test accuracies of $100\%$ and $95\%$ refer to separately evaluated selected parameters}
\label{fig:mnist_hardware}
\end{figure}

The eight-qubit QCNN has 63 trainable parameters and is optimized with OCP-RMSprop for ten mini batch updates. These updates are not ten complete passes through the dataset. With regularization $\epsilon=0.1$, RMSprop learning rate $0.05$, and learning-rate decay $0.01$, the simulator loss decreases from $0.295141$ to $0.171905$. Its lowest recorded loss is $0.165340$ at step 9. The hardware loss decreases from $0.263768$ to $0.181417$, which is also its minimum and occurs at step 10 (Fig.~\ref{fig:mnist_hardware}). The two terminal losses differ by approximately $0.009513$ MSE, although the paths are not monotone.

The selected simulator parameters achieve $92.5\%$ training accuracy and $100\%$ test accuracy; the hardware parameters achieve $86.25\%$ and $95\%$, respectively. The latter test result is 38/40 correct, compared with 40/40 in simulation. Consequently, the results concern this small binary-classification split rather than general MNIST performance.

The implemented OCP classifier evaluates an unshifted batch and four shifts per parameter, giving $16(1+4\times63)=4048$ sample-circuit expectation evaluations per update and 40,480 during training. The hardware session contains these ten training jobs and one post-training evaluation job with 200 additional sample expectations. This verifies the training cost for the displayed ten-step run; the 404,800 count in the 100 step benchmark does not apply to this figure. One selected OCP-RMSprop trajectory does not establish superiority over other estimators or update rules on hardware.

\subsection{Effect of the Update Rule for a Fixed Estimator}
\label{subsec:results_update_rules}

Table~\ref{tab:update_rule_effects} compares SGD, Adam, and RMSprop while holding the estimator fixed. Each entry is a mean across the three seeds. The range across update rules measures how much the reported outcome changes when the classical update is replaced; it does not describe variability across seeds. The learning rate schedule and other settings remain those of the benchmark. The results therefore compare the specified configurations, rather than the best performance attainable after separate tuning of each rule.

\begin{table*}[t]
\centering
\small
\caption{Update rule comparison for each fixed estimator, using means across seeds 123, 1234, and 42 at the reported 100 step checkpoint. MaxCut reports mean best cost, VQE reports mean absolute error in mHa, and classification reports mean test accuracy in percent. Range is the difference between the largest and smallest displayed means across update rules; it is not a measure of variation across seeds. Accuracy ranges are in percentage points. These comparisons retain the original stopping rules and are not matched by total measurement cost.}
\label{tab:update_rule_effects}
\setlength{\tabcolsep}{5pt}
\begin{tabular}{@{}llrrrrl@{}}
\toprule
\textbf{Task} & \textbf{Estimator} & \textbf{SGD} & \textbf{Adam} & \textbf{RMSprop} & \textbf{Range} & \textbf{Best rule} \\
\midrule
MaxCut & FiniteDiff & $-4.154948$ & $-4.216146$ & $-4.329427$ & $0.174479$ & RMSprop \\
MaxCut & SPSA & $-4.266927$ & $-4.024740$ & $-4.128906$ & $0.242187$ & SGD \\
MaxCut & PGPE & $-4.345052$ & $-4.182292$ & $-4.250000$ & $0.162760$ & SGD \\
MaxCut & PSR & $-4.220052$ & $-4.341146$ & $-4.303386$ & $0.121094$ & Adam \\
MaxCut & OCP & $-4.238281$ & $-4.109375$ & $-4.305989$ & $0.196614$ & RMSprop \\
\addlinespace
$H_2$ VQE & FiniteDiff & $4.1601$ & $3.9324$ & $2.6949$ & $1.4652$ & RMSprop \\
$H_2$ VQE & SPSA & $5.2845$ & $5.7028$ & $6.2846$ & $1.0001$ & SGD \\
$H_2$ VQE & PGPE & $5.8863$ & $6.8421$ & $4.4455$ & $2.3966$ & RMSprop \\
$H_2$ VQE & PSR & $4.3320$ & $7.7990$ & $6.5406$ & $3.4670$ & SGD \\
$H_2$ VQE & OCP & $4.2140$ & $7.5233$ & $6.9856$ & $3.3093$ & SGD \\
\addlinespace
Iris & FiniteDiff & $73.33$ & $95.00$ & $98.33$ & $25.00$ & RMSprop \\
Iris & SPSA & $75.00$ & $93.33$ & $91.67$ & $18.33$ & Adam \\
Iris & PGPE & $73.33$ & $90.00$ & $86.67$ & $16.67$ & Adam \\
Iris & PSR & $75.00$ & $95.00$ & $98.33$ & $23.33$ & RMSprop \\
Iris & OCP & $88.33$ & $96.67$ & $100.00$ & $11.67$ & RMSprop \\
\addlinespace
QCNN & FiniteDiff & $62.50$ & $91.67$ & $91.67$ & $29.17$ & Adam, RMSprop \\
QCNN & SPSA & $60.00$ & $86.67$ & $88.33$ & $28.33$ & RMSprop \\
QCNN & PGPE & $57.50$ & $92.50$ & $89.17$ & $35.00$ & Adam \\
QCNN & PSR & $60.83$ & $93.33$ & $92.50$ & $32.50$ & Adam \\
QCNN & OCP & $86.67$ & $95.00$ & $96.67$ & $10.00$ & RMSprop \\
\addlinespace
\bottomrule
\end{tabular}
\end{table*}

For MaxCut, the range in mean cost across update rules is between $0.121094$ for PSR and $0.242187$ for SPSA. Replacing Adam with SGD in SPSA reduces mean cost from $-4.024740$ to $-4.266927$. PGPE also performs best with SGD, whereas PSR performs best with Adam and finite differences and OCP perform best with RMSprop. The preferred update rule therefore depends on the estimator even when the graph, objective, and checkpoint are unchanged.

The same dependence is observed in VQE. With finite differences, RMSprop reduces mean absolute error from $4.1601$~mHa with SGD to $2.6949$~mHa, a reduction of $1.4652$~mHa. With PGPE, RMSprop gives $4.4455$~mHa compared with $6.8421$~mHa for Adam. Conversely, SGD gives the lowest mean error for SPSA, PSR, and OCP. The largest ranges occur for PSR ($3.4670$~mHa) and OCP ($3.3093$~mHa). Thus, the favorable average result of FiniteDiff-RMSprop does not imply that RMSprop improves every energy estimator.

For classification, the direction of the comparison is more consistent: both adaptive rules exceed SGD in mean test accuracy for every estimator on both tasks. On Iris, replacing SGD with the better adaptive rule increases mean test accuracy by $11.67$ to $25.00$ percentage points, depending on the estimator. For QCNN, the corresponding increase is $10.00$ to $35.00$ percentage points. The largest change is obtained with PGPE, from $57.50\%$ with SGD to $92.50\%$ with Adam. OCP is less sensitive to update choice, but RMSprop still improves its mean test accuracy over SGD by $11.67$ percentage points on Iris and $10.00$ percentage points on QCNN. These observations support adaptive updates for the reported classification configurations, while the energy tasks require a choice specific to the estimator.

The comparison is descriptive. Early stopping can change the total number of evaluations, and selecting the minimum observed training loss can favor sampling fluctuations or an easier batch. The classification averages also retain the original data partitions and preprocessing. Consequently, the differences quantify observed performance under the implemented protocol and do not isolate a general causal effect of the update rule.

\subsection{Objective Quality and Measurement Budgets}
\label{subsec:results_measurement_budgets}
\label{subsec:results_performance_cost}

A common number of optimization steps does not imply a common quantum-measurement budget. Under the protocol used in this study, each logical circuit is evaluated with 256 requested shots, so equal logical-circuit counts also correspond to equal requested training-shot counts. The resource comparisons reported here exclude post-training accuracy evaluation, hardware calibration, mitigation overhead, and other service-level execution costs. They therefore quantify logical training requests rather than complete hardware resource consumption.

For MaxCut, SPSA-SGD reaches a best sampled cost of $-4.359375$ using 300 logical circuit evaluations, corresponding to 76,800 requested shots. Under the same circuit count, SPSA-RMSprop reaches $-4.101562$. FiniteDiff-Adam reaches the same best sampled cost as SPSA-SGD, but requires 900 logical circuit evaluations. PGPE-RMSprop produces the lowest recorded seed-42 cost, $-4.363281$, using 1,100 logical circuits. Relative to SPSA-SGD, this improves the recorded objective by $0.003906$ while using approximately $3.67$ times as many logical circuit evaluations. The substantially larger circuit counts required by PSR and OCP do not yield a lower sampled minimum for this seed. These observations illustrate that additional measurement expenditure does not necessarily translate into a proportionate improvement in the recorded objective. Across the three seeds, PGPE-SGD attains the lowest mean best cost, whereas the seed-42 resource comparison highlights SPSA-SGD as a lower-cost configuration with a closely comparable recorded minimum. Configurations that terminate before the 100-step reporting point are evaluated using their actual recorded resource consumption; their trajectories are not extrapolated to a larger budget.

For Iris classification, FiniteDiff and PSR each require 52,800 logical circuit evaluations over 100 modular steps. At this matched circuit count, their RMSprop variants both attain $100\%$ test accuracy for seed 42. With Adam, PSR attains $90\%$ and FiniteDiff $85\%$, while with SGD the corresponding accuracies are $80\%$ and $75\%$. OCP-RMSprop also reaches $100\%$ test accuracy, but requires 104,000 logical circuit evaluations. Its main distinction is therefore not a higher seed-42 accuracy at lower cost, but the consistently high recorded accuracy across the three seed-specific train--test splits. COBYLA also reaches $100\%$ test accuracy for seed 42 using only 1,600 training circuit evaluations, whereas Bayesian optimization reaches $85\%$ at the same recorded cost. However, COBYLA's mean test accuracy across the three seeds is $81.67\%$. The inexpensive perfect result from one seed should therefore be interpreted as a single recorded outcome rather than as evidence of consistently superior performance.

The QCNN results show an even wider spread in accuracy at comparable circuit counts. SPSA-Adam and SPSA-RMSprop each reach $90\%$ test accuracy using 4,800 logical circuit evaluations, whereas SPSA-SGD reaches $47.5\%$ at the same cost. PGPE-Adam reaches $97.5\%$ using 17,600 circuits; PGPE-RMSprop reaches $90\%$, and PGPE-SGD reaches $47.5\%$, under the same recorded circuit count. At 203,200 logical circuit evaluations, shared by the FiniteDiff and PSR configurations, PSR-Adam records the highest seed-42 test accuracy at $97.5\%$, followed by FiniteDiff-RMSprop at $95\%$. OCP-Adam and OCP-RMSprop both reach $100\%$ test accuracy using 404,800 logical circuit evaluations. Compared with PGPE-Adam, this corresponds to one additional correct classification among 40 test examples while using approximately 23 times as many training circuit evaluations. This difference should be interpreted as an observed accuracy--cost trade-off, not as evidence that the full OCP budget is required to obtain $100\%$ test accuracy.

The recorded comparisons therefore identify configurations that attain different objective values at different resource levels, but they do not define a universal equal-budget ordering of the estimators. In particular, the available logs do not report OCP test accuracy after only the number of steps that could be executed within the PGPE circuit budget. Establishing such a comparison would require storing parameter checkpoints at matched cumulative circuit counts and evaluating the corresponding models using a common validation and test protocol. The QCNN derivative-free methods use changing mini batches of 16 samples per query. This batch size alone does not establish equivalent training-data exposure across optimizer configurations, whose numbers of objective queries and parameter perturbations differ.

For the $H_2$ VQE workload, the available resource records support comparison by objective-evaluation count but not by a verified logical-circuit or shot budget. The number of commuting measurement groups used for each energy evaluation was not retained, so objective evaluations cannot be converted reliably into total logical circuit executions or requested shots. At seed 42, the three FiniteDiff variants each use 700 energy evaluations. Their reported absolute errors relative to the FCI reference are $4.0632$~mHa for SGD, $4.2941$~mHa for Adam, and $0.7952$~mHa for RMSprop. These results demonstrate that substantially different sampled energy errors can occur at the same objective-evaluation count. They do not, however, establish an equal measurement-budget comparison among VQE estimators. Such a comparison would require the missing measurement-group records, and independent reevaluation of the selected parameter states would also be needed to distinguish persistent energy accuracy from favorable finite-shot samples.

The selected hardware runs follow the same accounting convention, with
training counts and session records in Appendix~\ref{sec:hardware_logs}.
Their workload-specific budgets do not establish an equal-cost comparison or
a simulator-to-hardware speed-up. Training time, submission latency, quantum
execution time, and calibration overhead remain distinct quantities.

\section{Discussion and Limitations}\label{sec:discussion}

The distinction between peak and average performance changes the preferred configuration for both objective-minimization tasks: SPSA-SGD and PSR-Adam share the MaxCut peak, whereas PGPE-SGD leads on average; PGPE-SGD gives the closest individual VQE estimate, whereas FiniteDiff-RMSprop minimizes mean absolute error. In classification, OCP-RMSprop combines a shared perfect single-seed result with the highest mean test accuracy on both tasks. Adaptive updates, particularly RMSprop, occupy the strongest average-performance positions in VQE and classification, but their benefit depends on the estimator and workload. Standard SGD supports exceptional peaks with SPSA and PGPE and also the best MaxCut average. Thus, neither the update rule nor the direction estimator alone determines performance. Selection should reflect the task metric, sensitivity across seeds, and evaluation cost, with peak outcomes and mean outcomes reported separately.

The comparisons support a limited but consistent observation: all four selected methods improve their monitored objective at some point on both backends, while the extent of improvement and subsequent fluctuation differ. The matched MaxCut terminal costs are relatively close, Iris has a larger terminal mini batch loss on hardware, and the QCNN ends with similar mini batch losses but different held-out accuracies. In the $H_2$ case, the hardware trajectory improves transiently and then deteriorates; its total energy remains substantially above the FCI reference. These observations do not establish a universal optimizer ranking or isolate hardware noise from compilation, mitigation, stochastic sampling, and calibration differences.

The distinction between terminal and best observed values matters in every selected task; in three of the hardware runs, the minimum occurs before the final recorded point. The hardware Iris and MaxCut minima occur before their final recorded points, and the H$_2$ hardware minimum at step 22 is followed by a substantial increase. Classifier test accuracies evaluate selected parameter vectors after training and must not be presented as the accuracy of the last plotted mini batch. For energy minimization, the best measured value can also be biased downward by selection under finite sampling; agreement with chemical accuracy requires an independent precision assessment rather than selection of a favorable point.

The fixed graph definition illustrates why reproducibility metadata must include the problem instance as well as the optimizer seed. Fixing the MaxCut graph separately prevents seed-dependent changes in the target objective. Similar care is needed for classifier splits and batch sampling: the QCNN test split contains only 40 samples and Iris contains 20. The hardware execution records verify execution settings and observations, but do not encode every classical training choice. In particular, the QCNN optimizer identity is supported by the experiment context and recorded parameter-shift structure, not by a Runtime optimizer-name field.

Each selected hardware case and its matched simulator comparison is represented by one trajectory, and hardware tasks were executed on different dates. These single runs do not support confidence intervals, isolation of hardware noise effects, or a hardware ranking across seeds. The tables at 100 steps across seeds establish the descriptive simulator rankings discussed above; the shorter hardware cases provide complementary evidence for selected configurations and do not establish those rankings on hardware. Repeated runs with fixed problem instances, independently controlled random streams, recorded calibration metadata, and independently reevaluated selected parameters are needed to quantify reproducibility and generalization.

\section{Conclusion}\label{sec:conclusion}

The modular framework separates search-direction estimation from classical update rules and makes their evaluation costs explicit. Across the three simulator seeds, PGPE-SGD attains the lowest mean MaxCut cost, FiniteDiff-RMSprop the lowest mean absolute VQE error, and OCP-RMSprop the highest mean test accuracy on both classifiers. The best individual outcomes do not always identify the strongest average configuration. Comparisons within a fixed estimator show that the update rule can substantially change accuracy and objective quality; the recorded resource counts permit comparisons within equal cost groups but do not constitute a common budget experiment across all methods. The present simulator and hardware comparisons use SPSA-SGD for MaxCut, COBYLA for Iris, OCP-RMSprop for QCNN, and PGPE-SGD for $H_2$ VQE, with verified archived job references and workload-specific budgets. Reporting terminal objectives, best observed objectives, and independently evaluated classifier accuracies separately avoids conflating different notions of performance. The resulting evidence describes the selected runs and their resource demands; it does not establish a universally superior optimizer or a statistically isolated hardware noise effect.
\section*{Statements and Declarations}

\textbf{Funding}

This work was partially supported by RExQTCS: Romanian Excellence Quantum Technologies enhancing Cybersecurity, a grant of the Ministry of Education and Research, CCCDI-UEFISCDI, project number PN-IV-P6-6.1-CoEx-2024-0214, within PNCDI IV.

\textbf{Competing Interests}

The authors have no relevant financial or non-financial interests to disclose.

\textbf{Ethics Approval and Consent to Participate}

Not applicable. This study did not involve human participants or animals.

\textbf{Consent for Publication}

Not applicable.

\textbf{Data Availability}

The Iris and MNIST datasets used in this study are publicly available datasets.
The preprocessing procedures and train--test split protocols are described in
Section~\ref{subsec:methods_data}.

\textbf{Materials Availability}

Not applicable.

\textbf{Code Availability} \\
The Python source code for all simulations and physical QPU experiments is available at \href{https://doi.org/10.5281/zenodo.23081874}{DOI: 10.5281/zenodo.23081874}.

\textbf{Author Contributions}

C.C.: Software, Investigation, Writing -- original draft.
S.B.: Conceptualization, Resources, Supervision, Writing -- review and editing.
A.A.: Resources, Supervision, Writing -- review and editing.
All authors reviewed and approved the final manuscript.
\bibliography{bibliography}

@article{McClean2018Barren,
  author  = {McClean, Jarrod R. and Boixo, Sergio and Smelyanskiy, Vadim N. and Babbush, Ryan and Neven, Hartmut},
  title   = {Barren plateaus in quantum neural network training landscapes},
  journal = {Nature Communications},
  volume  = {9},
  pages   = {4812},
  year    = {2018},
  doi     = {10.1038/s41467-018-07090-4}
}

@article{cerezo2021cost,
  author  = {Cerezo, M. and Sone, Akira and Volkoff, Tyler and Cincio, Lukasz and Coles, Patrick J.},
  title   = {Cost function dependent barren plateaus in shallow parametrized quantum circuits},
  journal = {Nature Communications}, 
  volume  = {12}, 
  number  = {1}, 
  pages   = {1791}, 
  year    = {2021},
  doi     = {10.1038/s41467-021-21728-w}
}

@article{Schuld2019Evaluating,
  author  = {Schuld, Maria and Bergholm, Ville and Gogolin, Christian and Izaac, Josh and Killoran, Nathan},
  title   = {Evaluating analytic gradients on quantum hardware},
  journal = {Physical Review A},
  volume  = {99},
  number  = {3},
  pages   = {032331},
  year    = {2019},
  doi     = {10.1103/PhysRevA.99.032331}
}

@article{Singh2023Benchmarking,
  author  = {Singh, Harshdeep and Mishra, Sabyashachi and Majumder, Sonjoy},
  title   = {Benchmarking of different optimizers in the variational quantum algorithms for applications in quantum chemistry},
  journal = {The Journal of Chemical Physics},
  volume  = {159},
  number  = {4},
  pages   = {044117},
  year    = {2023},
  doi     = {10.1063/5.0161057}
}

@article{sim2019expressibility,
  author  = {Sim, Sukin and Johnson, Peter D. and Aspuru-Guzik, Al{\'a}n},
  title   = {Expressibility and entangling capability of parameterized quantum circuits for hybrid quantum-classical algorithms},
  journal = {Advanced Quantum Technologies}, 
  volume  = {2}, 
  number  = {12}, 
  pages   = {1900070}, 
  year    = {2019},
  doi     = {10.1002/qute.201900070}
}

@article{peruzzo2014variational,
  author  = {Peruzzo, Alberto and McClean, Jarrod and Shadbolt, Peter and Yung, Man-Hong and Zhou, Xiao-Qi and Love, Peter J. and Aspuru-Guzik, Al{\'a}n and O'Brien, Jeremy L.},
  title   = {A variational eigenvalue solver on a photonic quantum processor},
  journal = {Nature Communications}, 
  volume  = {5}, 
  pages   = {4213}, 
  year    = {2014},
  doi     = {10.1038/ncomms5213}
}

@article{Stokes2020Quantum,
  author  = {Stokes, James and Izaac, Josh and Killoran, Nathan and Carleo, Giuseppe},
  title   = {Quantum natural gradient},
  journal = {Quantum},
  volume  = {4},
  pages   = {269},
  year    = {2020},
  doi     = {10.22331/q-2020-05-25-269}
}

@article{spall1992multivariate,
  author  = {Spall, James C.},
  title   = {Multivariate stochastic approximation using a simultaneous perturbation gradient approximation},
  journal = {IEEE Transactions on Automatic Control}, 
  volume  = {37}, 
  number  = {3}, 
  pages   = {332--341}, 
  year    = {1992},
  doi     = {10.1109/9.119632}
}

@article{tilly2022variational,
  author  = {Tilly, Jules and Chen, Hongxiang and Cao, Shuxiang and Picozzi, Dario and Setia, Kanav and Li, Ying and Grant, Edward and Wossnig, Leonard and Rungger, Ivan and Booth, George H. and Tennyson, Jonathan},
  title   = {The variational quantum eigensolver: A review of methods and best practices},
  journal = {Physics Reports}, 
  volume  = {986}, 
  pages   = {1--128}, 
  year    = {2022},
  doi     = {10.1016/j.physrep.2022.08.003}
}

@incollection{powell1994direct,
  author    = {Powell, Michael J. D.},
  title     = {A direct search optimization method that models the objective and constraint functions by linear interpolation},
  booktitle = {Advances in Optimization and Numerical Analysis}, 
  editor    = {Gomez, Suzanne and Hennart, Jean-Pierre},
  publisher = {Kluwer Academic Publishers}, 
  address   = {Boston}, 
  pages     = {51--67}, 
  year      = {1994}, 
  doi       = {10.1007/978-94-015-8330-5_4}
}

@inproceedings{Lavrijsen2020Classical,
  author    = {Lavrijsen, Wim and Tudor, Ana and M{\"u}ller, Juliane and Iancu, Costin and de Jong, Wibe},
  title     = {Classical optimizers for noisy intermediate-scale quantum devices},
  booktitle = {2020 IEEE International Conference on Quantum Computing and Engineering (QCE)},
  pages     = {267--277},
  year      = {2020},
  doi       = {10.1109/QCE49297.2020.00041},
}

@article{PellowJarman2021Comparison,
  author  = {Pellow-Jarman, Alexander and Sinayskiy, Ilya and Pillay, Anban and Petruccione, Francesco},
  title   = {A comparison of various classical optimizers for a variational quantum linear solver},
  journal = {Quantum Information Processing},
  volume  = {20},
  pages   = {202},
  year    = {2021},
  doi     = {10.1007/s11128-021-03140-x}
}

@article{BonetMonroig2023Performance,
  author  = {Bonet-Monroig, Xavier and Wang, Hao and Vermetten, Diederick and Senjean, Bruno and Moussa, Charles and B{\"a}ck, Thomas and Dunjko, Vedran and O'Brien, Thomas E.},
  title   = {Performance comparison of optimization methods on variational quantum algorithms},
  journal = {Physical Review A},
  volume  = {107},
  number  = {3},
  pages   = {032407},
  year    = {2023},
  doi     = {10.1103/PhysRevA.107.032407}
}

@article{Gacon2021SPSA,
  author  = {Gacon, Julien and Zoufal, Christa and Carleo, Giuseppe and Woerner, Stefan},
  title   = {Simultaneous perturbation stochastic approximation of the quantum {Fisher} information},
  journal = {Quantum},
  volume  = {5},
  pages   = {567},
  year    = {2021},
  doi     = {10.22331/q-2021-10-20-567}
}

@article{sweke2020stochastic,
  author  = {Sweke, Ryan and Bergholm, Ville and McArdle, Sam and Piroli, Lorenzo and Karalekas, James and Killoran, Nathan},
  title   = {Stochastic gradient descent for hybrid quantum-classical optimization},
  journal = {Quantum}, 
  volume  = {4}, 
  pages   = {314}, 
  year    = {2020},
  doi     = {10.22331/q-2020-08-31-314}
}

@article{kiefer1952stochastic,
  author  = {Kiefer, Jack and Wolfowitz, Jacob},
  title   = {Stochastic estimation of the maximum of a regression function},
  journal = {The Annals of Mathematical Statistics}, 
  volume  = {23}, 
  number  = {3}, 
  pages   = {462--466}, 
  year    = {1952}, 
  doi     = {10.1214/aoms/1177729392}
}

@article{sehnke2010parameter,
  author  = {Sehnke, Frank and Osendorfer, Christian and R{\"u}cksties, Thomas and Graves, Alex and Peters, Jan and Schmidhuber, J{\"u}rgen},
  title   = {Parameter-exploring policy gradients},
  journal = {Neural Networks}, 
  volume  = {23}, 
  number  = {4}, 
  pages   = {551--559}, 
  year    = {2010},
  doi     = {10.1016/j.neunet.2010.02.026}
}

@article{nelder1965simplex,
  author    = {Nelder, John A. and Mead, Roger},
  title     = {A simplex method for function minimization},
  journal   = {The Computer Journal},
  volume    = {7},
  number    = {4},
  pages     = {308--313},
  year      = {1965},
  publisher = {Oxford University Press},
  doi       = {10.1093/comjnl/7.4.308}
}

@article{frazier2018tutorial,
  author  = {Frazier, Peter I.},
  title   = {A tutorial on {Bayesian} optimization},
  journal = {arXiv preprint arXiv:1807.02811},
  year    = {2018},
  doi     = {10.48550/arXiv.1807.02811},
  url     = {https://doi.org/10.48550/arXiv.1807.02811}
}

@article{farhi2014quantum,
  author  = {Farhi, Edward and Goldstone, Jeffrey and Gutmann, Sam},
  title   = {A quantum approximate optimization algorithm},
  journal = {arXiv preprint arXiv:1411.4028},
  year    = {2014},
  doi     = {10.48550/arXiv.1411.4028},
  url     = {https://doi.org/10.48550/arXiv.1411.4028}
}

@article{cong2019quantum,
  author    = {Cong, Iris and Choi, Soonwon and Lukin, Mikhail D.},
  title     = {Quantum convolutional neural networks},
  journal   = {Nature Physics},
  volume    = {15},
  number    = {12},
  pages     = {1273--1278},
  year      = {2019},
  publisher = {Nature Publishing Group UK London},
  doi       = {10.1038/s41567-019-0648-8}
}

@article{Nannicini2019Performance,
  author  = {Nannicini, Giacomo},
  title   = {Performance of hybrid quantum-classical variational heuristics for combinatorial optimization},
  journal = {Physical Review E},
  volume  = {99},
  number  = {1},
  pages   = {013304},
  year    = {2019},
  doi     = {10.1103/PhysRevE.99.013304}
}

@article{Mari2021Estimating,
  author  = {Mari, Andrea and Bromley, Thomas R. and Killoran, Nathan},
  title   = {Estimating the gradient and higher-order derivatives on quantum hardware},
  journal = {Physical Review A},
  volume  = {103},
  number  = {1},
  pages   = {012405},
  year    = {2021},
  doi     = {10.1103/PhysRevA.103.012405}
}

@article{Kubler2020Adaptive,
  author  = {K{\"u}bler, Jonas M. and Arrasmith, Andrew and Cincio, Lukasz and Coles, Patrick J.},
  title   = {An adaptive optimizer for measurement-frugal variational algorithms},
  journal = {Quantum},
  volume  = {4},
  pages   = {263},
  year    = {2020},
  doi     = {10.22331/q-2020-05-11-263}
}

@article{Arrasmith2020Operator,
  author  = {Arrasmith, Andrew and Cincio, Lukasz and Somma, Rolando D. and Coles, Patrick J.},
  title   = {Operator sampling for shot-frugal optimization in variational algorithms},
  journal = {arXiv preprint arXiv:2004.06252},
  year    = {2020},
  doi     = {10.48550/arXiv.2004.06252},
  url     = {https://doi.org/10.48550/arXiv.2004.06252}
}

@article{Preskill2018NISQ,
  author  = {Preskill, John},
  title   = {Quantum computing in the {NISQ} era and beyond},
  journal = {Quantum},
  volume  = {2},
  pages   = {79},
  year    = {2018},
  doi     = {10.22331/q-2018-08-06-79}
}

@article{Cerezo2021Variational,
  author  = {Cerezo, M. and Arrasmith, Andrew and Babbush, Ryan and Benjamin, Simon C. and Endo, Suguru and Fujii, Keisuke and McClean, Jarrod R. and Mitarai, Kosuke and Yuan, Xiao and Cincio, Lukasz and Coles, Patrick J.},
  title   = {Variational quantum algorithms},
  journal = {Nature Reviews Physics},
  volume  = {3},
  pages   = {625--644},
  year    = {2021},
  doi     = {10.1038/s42254-021-00348-9}
}

@article{Bharti2022Noisy,
  author  = {Bharti, Kishor and Cervera-Lierta, Alba and Kyaw, Thi Ha and Haug, Tobias and Alperin-Lea, Sumner and Anand, Abhinav and Degroote, Matthias and Heimonen, Hermanni and Kottmann, Jakob S. and Menke, Tim and Mok, Wai-Keong and Sim, Sukin and Kwek, Leong-Chuan and Aspuru-Guzik, Al{\'a}n},
  title   = {Noisy intermediate-scale quantum algorithms},
  journal = {Reviews of Modern Physics},
  volume  = {94},
  number  = {1},
  pages   = {015004},
  year    = {2022},
  doi     = {10.1103/RevModPhys.94.015004}
}

@article{Wang2021NoiseInduced,
  author  = {Wang, Samson and Fontana, Enrico and Cerezo, M. and Sharma, Kunal and Sone, Akira and Cincio, Lukasz and Coles, Patrick J.},
  title   = {Noise-induced barren plateaus in variational quantum algorithms},
  journal = {Nature Communications},
  volume  = {12},
  pages   = {6961},
  year    = {2021},
  doi     = {10.1038/s41467-021-27045-6}
}

@inproceedings{Kingma2015Adam,
  author    = {Kingma, Diederik P. and Ba, Jimmy},
  title     = {{Adam}: A method for stochastic optimization},
  booktitle = {Proceedings of the 3rd International Conference on Learning Representations ({ICLR})},
  year      = {2015},
  eprint={1412.6980},
  archivePrefix={arXiv},
  primaryClass={cs.LG},
  url={https://arxiv.org/abs/1412.6980}, 
}

@misc{Tieleman2012RMSProp,
  author       = {Tieleman, Tijmen and Hinton, Geoffrey},
  title        = {Lecture 6.5---{RMSProp}: Divide the gradient by a running average of its recent magnitude},
  howpublished = {COURSERA: Neural Networks for Machine Learning},
  year         = {2012}
}

@article{Robbins1951Stochastic,
  author  = {Robbins, Herbert and Monro, Sutton},
  title   = {A stochastic approximation method},
  journal = {The Annals of Mathematical Statistics},
  volume  = {22},
  number  = {3},
  pages   = {400--407},
  year    = {1951},
  doi     = {10.1214/aoms/1177729586}
}

\begin{appendices}

\makeatletter
\@removefromreset{table}{section}
\@removefromreset{figure}{section}
\makeatother
\renewcommand{\thetable}{\arabic{table}}
\renewcommand{\thefigure}{\arabic{figure}}
\setcounter{table}{3}
\setcounter{figure}{6}

\section{Problem Hamiltonians} \label{secA1}
For $H_2$ at bond length $0.7414$ \AA\ in STO-3G, the Jordan--Wigner electronic Hamiltonian generated by the implemented PySCF/Qiskit Nature pipeline is
\begin{equation}
 \hat H_{\rm el}=\sum_P c_P P,\qquad \hat H_{\text{VQE}}=\hat H_{\rm el}+E_{\rm nuc}I. \label{eq:appendix_vqe_ham}
\end{equation}
Table~\ref{tab:h2_pauli_coefficients} lists the Hamiltonian coefficients generated for this problem. Pauli strings are written in Qiskit order $q_3q_2q_1q_0$; for example, \texttt{IIIZ} means $Z_0$. Coefficients are in Hartree and rounded to ten decimal places.
\begin{table}[ht]
\centering
\caption{Electronic Hamiltonian coefficients for $H_2$ at bond length 0.7414 \AA\ in the STO-3G basis}
\label{tab:h2_pauli_coefficients}
\begin{tabular}{lr}
\toprule
\textbf{Pauli string} & \textbf{Coefficient (Ha)} \\
\midrule
\texttt{IIII} & $-0.8126179630$ \\
\texttt{IIIZ} & $0.1711977490$ \\
\texttt{IIZI} & $-0.2227859304$ \\
\texttt{IIZZ} & $0.1205448221$ \\
\texttt{IZII} & $0.1711977490$ \\
\texttt{IZIZ} & $0.1686221916$ \\
\texttt{ZIII} & $-0.2227859304$ \\
\texttt{ZIIZ} & $0.1658670241$ \\
\texttt{YYYY} & $0.0453222021$ \\
\texttt{XXYY} & $0.0453222021$ \\
\texttt{YYXX} & $0.0453222021$ \\
\texttt{XXXX} & $0.0453222021$ \\
\texttt{IZZI} & $0.1658670241$ \\
\texttt{ZIZI} & $0.1743484419$ \\
\texttt{ZZII} & $0.1205448221$ \\
\bottomrule
\end{tabular}
\end{table}
The four off-diagonal terms are \texttt{XXXX}, \texttt{YYYY}, \texttt{XXYY}, and \texttt{YYXX}, with equal positive coefficients. The nuclear-repulsion contribution is $E_{\rm nuc}=0.7137539937$ Ha. Diagonalizing the electronic Hamiltonian and adding this constant gives $E_0=-1.1372701747$ Ha. The VQE results report this total molecular energy, not electronic energy alone.

\subsection{Fixed MaxCut graph generated with seed 42}
\label{subsec:appendix_graph42}
\begin{figure}[htbp]
\centering
\includegraphics[width=84mm]{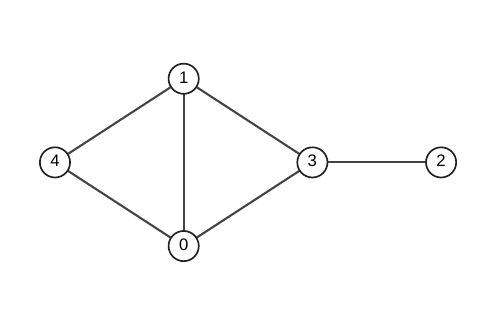}
\caption{MaxCut instance generated with seed 42 and used in every reported MaxCut run}
\label{fig:maxcut_instance}
\end{figure}

Figure~\ref{fig:maxcut_instance} shows the fixed graph defined in
Section~\ref{subsec:experimental_workloads}.
The maximum cut is five, attained by the partition $\{0,1,2\}\,|\,\{3,4\}$. The minimum expected negative cut is therefore $-5$. Optimizer seeds change initialization and sampling without changing this graph.

\section{Classical Optimizer Formulations} \label{secB1}
Adam maintains first and second moments of the supplied direction (including preconditioning when used). For iterations $k=0,1,\ldots$, moments start at $\mathbf m_{-1}=\mathbf v_{-1}=0$:
\begin{enumerate}
    \item First-Moment Tracking: The optimizer tracks the exponentially decaying average of past supplied directions (momentum vector, $\mathbf{m}_k$):
    \begin{equation}
        \mathbf{m}_k = \beta_1 \mathbf{m}_{k-1} + (1 - \beta_1) \hat{\mathbf{d}}^{(k)} \label{eq:adam_first_moment}
    \end{equation}
    where $\beta_1 = 0.9$ represents the momentum decay constant.
    \item Second-Moment Tracking: The optimizer tracks the exponentially decaying average of past squared supplied directions (second moment vector, $\mathbf{v}_k$):
    \begin{equation}
        \mathbf{v}_k = \beta_2 \mathbf{v}_{k-1} + (1 - \beta_2) \left( \hat{\mathbf{d}}^{(k)} \odot \hat{\mathbf{d}}^{(k)} \right) \label{eq:adam_second_moment}
    \end{equation}
    where $\beta_2 = 0.999$ represents the second moment decay constant, and $\odot$ denotes the element-wise Hadamard product.
    \item Bias Correction: Because the moments are initialized to zero, they are biased toward zero at early iterations. To correct this, bias-corrected estimators are computed:
    \begin{equation}
        \hat{\mathbf{m}}_k = \frac{\mathbf{m}_k}{1 - \beta_1^{k+1}}, \quad \hat{\mathbf{v}}_k = \frac{\mathbf{v}_k}{1 - \beta_2^{k+1}} \label{eq:bias_correction}
    \end{equation}
    \item Parameter Update Step: The continuous parameters are adjusted using the adaptive scale:
    \begin{equation}
        \boldsymbol{\theta}^{(k+1)} = \boldsymbol{\theta}^{(k)} - \frac{\eta_k \hat{\mathbf{m}}_k}{\sqrt{\hat{\mathbf{v}}_k} + \epsilon} \label{eq:parameter_update_step}
    \end{equation}
    where $\epsilon = 10^{-8}$ ensures numerical stability, and the learning rate follows Eq.~\eqref{eq:lr_decay} in Section~\ref{subsec:experimental_hyperparameters}.
\end{enumerate}
RMSprop uses its own uncorrected second-moment recurrence, with $\rho=0.9$, $\mathbf v_{-1}=0$, and $\epsilon=10^{-8}$:
\begin{align}
 \mathbf v_k&=\rho\mathbf v_{k-1}+(1-\rho)(\hat{\mathbf d}^{(k)}\odot\hat{\mathbf d}^{(k)}),\\
 \boldsymbol\theta^{(k+1)}&=\boldsymbol\theta^{(k)}-\eta_k\frac{\hat{\mathbf d}^{(k)}}{\sqrt{\mathbf v_k}+\epsilon}. \label{eq:rmsprop_update}
\end{align}
All vector divisions are componentwise. RMSprop has neither first-moment momentum nor bias correction in this implementation. SGD applies $\boldsymbol\theta^{(k+1)}=\boldsymbol\theta^{(k)}-\eta_k\hat{\mathbf d}^{(k)}$ directly.

\section{QPU Hardware Calibrations and Native Gate Compilation} \label{secC1}
The hardware experiments use the \texttt{ibm\_aachen} superconducting processor with an IBM Heron r3 architecture. The following calibration snapshot is linked to the reported hardware runs and summarizes the device characteristics recorded for those experiments:
\begin{itemize}
    \item Thermal Relaxation Coherence Time ($T_1$): Median value of $232.18$ $\mu\text{s}$
    \item Dephasing Coherence Time ($T_2$): Median value of $250.72$ $\mu\text{s}$
    \item Single-Qubit Gate Infidelity: Median error rate of $1.901 \times 10^{-4}$
    \item Two-Qubit Gate Infidelity (Native CZ): Median Controlled-Z ($\text{CZ}$) error rate of $1.596 \times 10^{-3}$
    \item Measurement Readout Error: Median readout error rate of $7.568 \times 10^{-3}$
\end{itemize}
The native two-qubit entangling gate on the Heron architecture is Controlled-Z ($\text{CZ}$). CNOT gates in the HEA, QAOA, and QCNN circuits are compiled into $\text{CZ}$ gates according to
\begin{equation}
    \text{CNOT}_{(i \rightarrow j)} = (I \otimes H) \cdot \text{CZ}_{ij} \cdot (I \otimes H) \label{eq:cnot_cz_compilation}
\end{equation}
This identity illustrates a possible CNOT decomposition. The actual native-gate counts and their contribution to error depend on transpilation and the selected hardware layout.

\section{Experimental Benchmark Data}\label{secD1}
The tables are grouped by task. For seed 42, performance is reported at the 100 step reporting point, followed by resource use and the recorded small change diagnostic. Seeds 123 and 1234 are represented by their 100 step results. The averages use all three seeds. The resource tables report the terminal objective and a retrospective small change diagnostic, which does not establish convergence. Reporting points refer to modular optimization steps; derivative-free methods retain their native query and stopping limits. All MaxCut runs use the same graph generated with seed 42.

\subsection{MaxCut QAOA benchmark tables}\label{subsec:appendix_maxcut}

\begin{table}[htbp]
\caption{Best observed cost and relative error for the MaxCut QAOA task in 256-shot simulation (100 step modular reporting point and native derivative-free limits, seed 42). The exact target is $-5.0$}
\label{tab:maxcut_100}
\centering
\setlength{\tabcolsep}{4pt}
\begin{tabular*}{\textwidth}{@{\extracolsep{\fill}}clccc@{}}
\toprule
\textbf{Rank} & \textbf{Algorithm} & \textbf{Best Cost} & \textbf{Dist. to Target} & \textbf{Relative Error (\%)} \\
\midrule
1 & PGPE-RMSprop & $-4.363281$ & $0.636719$ & $12.73$ \\
2 & FiniteDiff-Adam & $-4.359375$ & $0.640625$ & $12.81$ \\
3 & SPSA-SGD & $-4.359375$ & $0.640625$ & $12.81$ \\
4 & PGPE-SGD & $-4.359375$ & $0.640625$ & $12.81$ \\
5 & PGPE-Adam & $-4.355469$ & $0.644531$ & $12.89$ \\
6 & PSR-SGD & $-4.355469$ & $0.644531$ & $12.89$ \\
7 & FiniteDiff-RMSprop & $-4.339844$ & $0.660156$ & $13.20$ \\
8 & PSR-RMSprop & $-4.308594$ & $0.691406$ & $13.83$ \\
9 & OCP-RMSprop & $-4.308594$ & $0.691406$ & $13.83$ \\
10 & PSR-Adam & $-4.300781$ & $0.699219$ & $13.98$ \\
11 & OCP-SGD & $-4.214844$ & $0.785156$ & $15.70$ \\
12 & Bayesian & $-4.210938$ & $0.789062$ & $15.78$ \\
13 & OCP-Adam & $-4.140625$ & $0.859375$ & $17.19$ \\
14 & SPSA-RMSprop & $-4.101562$ & $0.898438$ & $17.97$ \\
15 & COBYLA & $-4.046875$ & $0.953125$ & $19.06$ \\
16 & Nelder-Mead & $-4.042969$ & $0.957031$ & $19.14$ \\
17 & FiniteDiff-SGD & $-4.031250$ & $0.968750$ & $19.38$ \\
18 & SPSA-Adam & $-4.015625$ & $0.984375$ & $19.69$ \\
\bottomrule
\end{tabular*}
\end{table}

\begin{sidewaystable}[p]
\caption{Resource utilization for the MaxCut QAOA task in 256-shot simulation at the 100 step modular reporting point and native derivative-free limits, seed 42. ``Steps/Queries'' denotes optimization steps for modular methods and objective queries for derivative-free methods. The final column reports the first iteration at which the absolute change between two consecutive recorded objective values was below $10^{-4}$; it is a retrospective trajectory diagnostic and not a convergence criterion.}
\label{tab:maxcut_resources}
\centering
\setlength{\tabcolsep}{3pt}
\begin{small}
\begin{tabular*}{\textheight}{@{\extracolsep{\fill}}lccccccc@{}}
\toprule
\textbf{Optimizer} & \textbf{Final Cost} & \textbf{Steps/Queries} & \makecell{\textbf{Objective}\\\textbf{Evals}} & \makecell{\textbf{Logical}\\\textbf{Circuits}} & \makecell{\textbf{Requested}\\\textbf{Shots}} & \textbf{Time (s)} & \makecell{\textbf{First Small-Change}\\\textbf{Iteration}} \\
\midrule
COBYLA & $-4.046875$ & $42$ & $42$ & $42$ & $10{,}752$ & $2.07$ & $26$ \\
Nelder-Mead & $-4.042969$ & $121$ & $121$ & $121$ & $30{,}976$ & $5.16$ & $43$ \\
Bayesian & $-4.210938$ & $100$ & $100$ & $100$ & $25{,}600$ & $127.33$ & $1$ \\
FiniteDiff-SGD & $-4.023438$ & $28$ & $252$ & $252$ & $64{,}512$ & $10.79$ & $2$ \\
FiniteDiff-Adam & $-4.332031$ & $100$ & $900$ & $900$ & $230{,}400$ & $25.45$ & $2$ \\
FiniteDiff-RMSprop & $-4.296875$ & $100$ & $900$ & $900$ & $230{,}400$ & $21.16$ & $16$ \\
SPSA-SGD & $-4.351562$ & $100$ & $300$ & $300$ & $76{,}800$ & $13.81$ & $46$ \\
SPSA-Adam & $-4.007812$ & $63$ & $189$ & $189$ & $48{,}384$ & $7.63$ & $11$ \\
SPSA-RMSprop & $-4.046875$ & $100$ & $300$ & $300$ & $76{,}800$ & $8.71$ & $14$ \\
PGPE-SGD & $-4.316406$ & $100$ & $1{,}100$ & $1{,}100$ & $281{,}600$ & $32.17$ & $11$ \\
PGPE-Adam & $-4.328125$ & $100$ & $1{,}100$ & $1{,}100$ & $281{,}600$ & $31.96$ & $3$ \\
PGPE-RMSprop & $-4.343750$ & $100$ & $1{,}100$ & $1{,}100$ & $281{,}600$ & $33.20$ & $8$ \\
PSR-SGD & $-4.347656$ & $100$ & $4{,}500$ & $4{,}500$ & $1{,}152{,}000$ & $97.88$ & $2$ \\
PSR-Adam & $-4.296875$ & $100$ & $4{,}500$ & $4{,}500$ & $1{,}152{,}000$ & $97.63$ & $18$ \\
PSR-RMSprop & $-4.300781$ & $100$ & $4{,}500$ & $4{,}500$ & $1{,}152{,}000$ & $98.22$ & $12$ \\
OCP-SGD & $-4.183594$ & $100$ & $28{,}900$ & $28{,}900$ & $7{,}398{,}400$ & $446.14$ & $23$ \\
OCP-Adam & $-4.136719$ & $100$ & $28{,}900$ & $28{,}900$ & $7{,}398{,}400$ & $475.92$ & $3$ \\
OCP-RMSprop & $-4.296875$ & $100$ & $28{,}900$ & $28{,}900$ & $7{,}398{,}400$ & $422.46$ & $22$ \\
\bottomrule
\end{tabular*}
\end{small}
\end{sidewaystable}

\begin{table}[htbp]
\caption{Best observed cost and relative error for the MaxCut QAOA task in 256-shot simulation (100 step modular reporting point, Seed 123). The exact target is $-5.0$. derivative-free results follow their native query and stopping limits; the reporting point does not imply an equal query budget.}
\label{tab:maxcut_100_seed123}
\centering
\setlength{\tabcolsep}{4pt}
\begin{tabular*}{\textwidth}{@{\extracolsep{\fill}}clccc@{}}
\toprule
\textbf{Rank} & \textbf{Optimizer} & \textbf{Best Cost} & \textbf{Dist. to Target} & \textbf{Relative Error (\%)} \\
\midrule
1 & SPSA-SGD & $-4.378906$ & $0.621094$ & $12.42$ \\
2 & PSR-Adam & $-4.378906$ & $0.621094$ & $12.42$ \\
3 & FiniteDiff-RMSprop & $-4.375000$ & $0.625000$ & $12.50$ \\
4 & SPSA-RMSprop & $-4.375000$ & $0.625000$ & $12.50$ \\
5 & FiniteDiff-SGD & $-4.367188$ & $0.632812$ & $12.66$ \\
6 & PGPE-SGD & $-4.359375$ & $0.640625$ & $12.81$ \\
7 & OCP-RMSprop & $-4.289062$ & $0.710938$ & $14.22$ \\
8 & PSR-RMSprop & $-4.277344$ & $0.722656$ & $14.45$ \\
9 & OCP-SGD & $-4.218750$ & $0.781250$ & $15.62$ \\
10 & SPSA-Adam & $-4.152344$ & $0.847656$ & $16.95$ \\
11 & PGPE-RMSprop & $-4.101562$ & $0.898438$ & $17.97$ \\
12 & PGPE-Adam & $-4.011719$ & $0.988281$ & $19.77$ \\
13 & FiniteDiff-Adam & $-4.000000$ & $1.000000$ & $20.00$ \\
14 & PSR-SGD & $-3.988281$ & $1.011719$ & $20.23$ \\
15 & Nelder-Mead & $-3.964844$ & $1.035156$ & $20.70$ \\
16 & OCP-Adam & $-3.925781$ & $1.074219$ & $21.48$ \\
17 & Bayesian & $-3.847656$ & $1.152344$ & $23.05$ \\
18 & COBYLA & $-3.808594$ & $1.191406$ & $23.83$ \\
\bottomrule
\end{tabular*}
\end{table}

\begin{table}[htbp]
\caption{Best observed cost and relative error for the MaxCut QAOA task in 256-shot simulation (100 step modular reporting point, Seed 1234). The exact target is $-5.0$. derivative-free results follow their native query and stopping limits; the reporting point does not imply an equal query budget.}
\label{tab:maxcut_100_seed1234}
\centering
\setlength{\tabcolsep}{4pt}
\begin{tabular*}{\textwidth}{@{\extracolsep{\fill}}clccc@{}}
\toprule
\textbf{Rank} & \textbf{Optimizer} & \textbf{Best Cost} & \textbf{Dist. to Target} & \textbf{Relative Error (\%)} \\
\midrule
1 & PSR-Adam & $-4.343750$ & $0.656250$ & $13.12$ \\
2 & PSR-RMSprop & $-4.324219$ & $0.675781$ & $13.52$ \\
3 & OCP-RMSprop & $-4.320312$ & $0.679688$ & $13.59$ \\
4 & PGPE-SGD & $-4.316406$ & $0.683594$ & $13.67$ \\
5 & PSR-SGD & $-4.316406$ & $0.683594$ & $13.67$ \\
6 & FiniteDiff-Adam & $-4.289062$ & $0.710938$ & $14.22$ \\
7 & PGPE-RMSprop & $-4.285156$ & $0.714844$ & $14.30$ \\
8 & OCP-SGD & $-4.281250$ & $0.718750$ & $14.37$ \\
9 & FiniteDiff-RMSprop & $-4.273438$ & $0.726562$ & $14.53$ \\
10 & OCP-Adam & $-4.261719$ & $0.738281$ & $14.77$ \\
11 & PGPE-Adam & $-4.179688$ & $0.820312$ & $16.41$ \\
12 & COBYLA & $-4.066406$ & $0.933594$ & $18.67$ \\
13 & FiniteDiff-SGD & $-4.066406$ & $0.933594$ & $18.67$ \\
14 & SPSA-SGD & $-4.062500$ & $0.937500$ & $18.75$ \\
15 & Nelder-Mead & $-3.996094$ & $1.003906$ & $20.08$ \\
16 & SPSA-RMSprop & $-3.910156$ & $1.089844$ & $21.80$ \\
17 & SPSA-Adam & $-3.906250$ & $1.093750$ & $21.88$ \\
18 & Bayesian & $-3.773438$ & $1.226562$ & $24.53$ \\
\bottomrule
\end{tabular*}
\end{table}

\begin{table}[htbp]
\caption{Averaged evaluation of optimizers on the MaxCut QAOA task across three random seeds (123, 1234, 42) at the 100 step modular reporting point. derivative-free results follow their native query and stopping limits; the reporting point does not imply an equal query budget.}
\label{tab:maxcut_average}
\centering
\setlength{\tabcolsep}{4pt}
\begin{tabular*}{\textwidth}{@{\extracolsep{\fill}}clccc@{}}
\toprule
\textbf{Rank} & \textbf{Optimizer} & \textbf{Avg. Best Cost} & \textbf{Avg. Dist. to Target} & \makecell{\textbf{Avg. Relative}\\\textbf{Error (\%)}} \\
\midrule
1 & PGPE-SGD & $-4.345052$ & $0.654948$ & $13.10$ \\
2 & PSR-Adam & $-4.341146$ & $0.658854$ & $13.18$ \\
3 & FiniteDiff-RMSprop & $-4.329427$ & $0.670573$ & $13.41$ \\
4 & OCP-RMSprop & $-4.305989$ & $0.694011$ & $13.88$ \\
5 & PSR-RMSprop & $-4.303386$ & $0.696614$ & $13.93$ \\
6 & SPSA-SGD & $-4.266927$ & $0.733073$ & $14.66$ \\
7 & PGPE-RMSprop & $-4.250000$ & $0.750000$ & $15.00$ \\
8 & OCP-SGD & $-4.238281$ & $0.761719$ & $15.23$ \\
9 & PSR-SGD & $-4.220052$ & $0.779948$ & $15.60$ \\
10 & FiniteDiff-Adam & $-4.216146$ & $0.783854$ & $15.68$ \\
11 & PGPE-Adam & $-4.182292$ & $0.817708$ & $16.35$ \\
12 & FiniteDiff-SGD & $-4.154948$ & $0.845052$ & $16.90$ \\
13 & SPSA-RMSprop & $-4.128906$ & $0.871094$ & $17.42$ \\
14 & OCP-Adam & $-4.109375$ & $0.890625$ & $17.81$ \\
15 & SPSA-Adam & $-4.024740$ & $0.975260$ & $19.51$ \\
16 & Nelder-Mead & $-4.001302$ & $0.998698$ & $19.97$ \\
17 & COBYLA & $-3.973958$ & $1.026042$ & $20.52$ \\
18 & Bayesian & $-3.944011$ & $1.055989$ & $21.12$ \\
\bottomrule
\end{tabular*}
\end{table}

\FloatBarrier

\subsection{Iris classification benchmark tables}\label{subsec:appendix_iris}

\begin{table}[htbp]
\caption{Best observed loss and accuracy metrics for the four-qubit Iris classification task in 256-shot simulation (100 step modular reporting point and native derivative-free limits, seed 42)}
\label{tab:iris_100_top}
\centering
\setlength{\tabcolsep}{4pt}
\begin{tabular*}{\textwidth}{@{\extracolsep{\fill}}clccc@{}}
\toprule
\textbf{Rank} & \textbf{Optimizer} & \textbf{Loss} & \makecell{\textbf{Train}\\\textbf{Accuracy (\%)}} & \makecell{\textbf{Test}\\\textbf{Accuracy (\%)}} \\
\midrule
1 & COBYLA & $0.037184$ & $100.00$ & $100.00$ \\
2 & FiniteDiff-RMSprop & $0.015455$ & $98.75$ & $100.00$ \\
3 & PSR-RMSprop & $0.011665$ & $98.75$ & $100.00$ \\
4 & OCP-RMSprop & $0.012451$ & $98.75$ & $100.00$ \\
5 & PGPE-Adam & $0.009300$ & $96.25$ & $90.00$ \\
6 & PGPE-RMSprop & $0.009272$ & $96.25$ & $90.00$ \\
7 & PSR-Adam & $0.010777$ & $98.75$ & $90.00$ \\
8 & OCP-Adam & $0.011987$ & $97.50$ & $90.00$ \\
9 & Bayesian & $0.078629$ & $95.00$ & $85.00$ \\
10 & FiniteDiff-Adam & $0.013467$ & $96.25$ & $85.00$ \\
11 & SPSA-Adam & $0.051228$ & $93.75$ & $85.00$ \\
12 & SPSA-RMSprop & $0.026241$ & $96.25$ & $85.00$ \\
13 & OCP-SGD & $0.033379$ & $97.50$ & $85.00$ \\
14 & PSR-SGD & $0.138197$ & $73.75$ & $80.00$ \\
15 & FiniteDiff-SGD & $0.142104$ & $71.25$ & $75.00$ \\
16 & SPSA-SGD & $0.156346$ & $63.75$ & $75.00$ \\
17 & PGPE-SGD & $0.114960$ & $80.00$ & $75.00$ \\
18 & Nelder-Mead & $0.186970$ & $61.25$ & $55.00$ \\
\bottomrule
\end{tabular*}
\end{table}

\begin{sidewaystable}[p]
\caption{Resource utilization for the four-qubit Iris classification task in 256-shot simulation at the 100 step modular reporting point and native derivative-free limits, seed 42. ``Steps/Queries'' denotes optimization steps for modular methods and objective queries for derivative-free methods. The final column reports the first iteration at which the absolute change between two consecutive recorded objective values was below $10^{-4}$; it is a retrospective trajectory diagnostic and not a convergence criterion.}
\label{tab:iris_100}
\centering
\setlength{\tabcolsep}{4pt}
\begin{small}
\begin{tabular*}{\textheight}{@{\extracolsep{\fill}}lccccccc@{}}
\toprule
\textbf{Optimizer} & \textbf{Final Loss} & \textbf{Steps/Queries} & \makecell{\textbf{Objective}\\\textbf{Evals}} & \makecell{\textbf{Logical}\\\textbf{Circuits}} & \makecell{\textbf{Requested}\\\textbf{Shots}} & \textbf{Time (s)} & \makecell{\textbf{First Small-Change}\\\textbf{Iteration}} \\
\midrule
COBYLA & $0.052316$ & $100$ & $100$ & $1{,}600$ & $409{,}600$ & $40.39$ & $41$ \\
Nelder-Mead & $0.217056$ & $279$ & $279$ & $4{,}464$ & $1{,}142{,}784$ & $249.06$ & $11$ \\
Bayesian & $0.078629$ & $100$ & $100$ & $1{,}600$ & $409{,}600$ & $416.24$ & $2$ \\
FiniteDiff-SGD & $0.187281$ & $100$ & $3{,}300$ & $52{,}800$ & $13{,}516{,}800$ & $929.80$ & $4$ \\
FiniteDiff-Adam & $0.044420$ & $100$ & $3{,}300$ & $52{,}800$ & $13{,}516{,}800$ & $1092.85$ & $37$ \\
FiniteDiff-RMSprop & $0.027587$ & $100$ & $3{,}300$ & $52{,}800$ & $13{,}516{,}800$ & $1066.26$ & $11$ \\
SPSA-SGD & $0.200388$ & $100$ & $300$ & $4{,}800$ & $1{,}228{,}800$ & $127.22$ & -- \\
SPSA-Adam & $0.090261$ & $100$ & $300$ & $4{,}800$ & $1{,}228{,}800$ & $125.31$ & $53$ \\
SPSA-RMSprop & $0.074401$ & $100$ & $300$ & $4{,}800$ & $1{,}228{,}800$ & $125.24$ & $1$ \\
PGPE-SGD & $0.126043$ & $100$ & $1{,}100$ & $17{,}600$ & $4{,}505{,}600$ & $404.74$ & $43$ \\
PGPE-Adam & $0.020151$ & $100$ & $1{,}100$ & $17{,}600$ & $4{,}505{,}600$ & $413.95$ & $64$ \\
PGPE-RMSprop & $0.017489$ & $100$ & $1{,}100$ & $17{,}600$ & $4{,}505{,}600$ & $393.74$ & $29$ \\
PSR-SGD & $0.184473$ & $100$ & $3{,}300$ & $52{,}800$ & $13{,}516{,}800$ & $1018.64$ & $14$ \\
OCP-SGD & $0.059799$ & $100$ & $6{,}500$ & $104{,}000$ & $26{,}624{,}000$ & $1874.01$ & $44$ \\
PSR-Adam & $0.030990$ & $100$ & $3{,}300$ & $52{,}800$ & $13{,}516{,}800$ & $896.38$ & $7$ \\
OCP-Adam & $0.036266$ & $100$ & $6{,}500$ & $104{,}000$ & $26{,}624{,}000$ & $1680.90$ & $60$ \\
PSR-RMSprop & $0.023490$ & $100$ & $3{,}300$ & $52{,}800$ & $13{,}516{,}800$ & $854.72$ & $3$ \\
OCP-RMSprop & $0.025423$ & $100$ & $6{,}500$ & $104{,}000$ & $26{,}624{,}000$ & $1662.28$ & $37$ \\
\bottomrule
\end{tabular*}
\end{small}
\end{sidewaystable}

\begin{table}[htbp]
\caption{Best observed loss and accuracy metrics for the 4-qubit Iris classification task in 256-shot simulation (100 step modular reporting point, Seed 123). derivative-free results follow their native query and stopping limits; the reporting point does not imply an equal query budget.}
\label{tab:iris_100_seed123}
\centering
\setlength{\tabcolsep}{6pt}
\begin{tabular*}{\textwidth}{@{\extracolsep{\fill}}clccc@{}}
\toprule
\textbf{Rank} & \textbf{Optimizer} & \textbf{Best Loss} & \makecell{\textbf{Train}\\\textbf{Accuracy (\%)}} & \makecell{\textbf{Test}\\\textbf{Accuracy (\%)}} \\
\midrule
1 & FiniteDiff-Adam & $0.090705$ & $98.75$ & $100.00$ \\
2 & SPSA-Adam & $0.114372$ & $98.75$ & $100.00$ \\
3 & OCP-Adam & $0.080906$ & $100.00$ & $100.00$ \\
4 & OCP-RMSprop & $0.063342$ & $98.75$ & $100.00$ \\
5 & FiniteDiff-RMSprop & $0.072222$ & $98.75$ & $95.00$ \\
6 & SPSA-RMSprop & $0.092929$ & $98.75$ & $95.00$ \\
7 & PSR-Adam & $0.067430$ & $98.75$ & $95.00$ \\
8 & PSR-RMSprop & $0.065356$ & $98.75$ & $95.00$ \\
9 & OCP-SGD & $0.110806$ & $95.00$ & $90.00$ \\
10 & SPSA-SGD & $0.170575$ & $52.50$ & $80.00$ \\
11 & PGPE-Adam & $0.096120$ & $90.00$ & $80.00$ \\
12 & FiniteDiff-SGD & $0.152019$ & $57.50$ & $75.00$ \\
13 & PGPE-SGD & $0.131916$ & $60.00$ & $75.00$ \\
14 & PSR-SGD & $0.153684$ & $60.00$ & $75.00$ \\
15 & Nelder-Mead & $0.167499$ & $50.00$ & $70.00$ \\
16 & PGPE-RMSprop & $0.081245$ & $81.25$ & $70.00$ \\
17 & COBYLA & $0.168144$ & $50.00$ & $65.00$ \\
18 & Bayesian & $0.134981$ & $50.00$ & $50.00$ \\
\bottomrule
\end{tabular*}
\end{table}

\begin{table}[htbp]
\caption{Best observed loss and accuracy metrics for the 4-qubit Iris classification task in 256-shot simulation (100 step modular reporting point, Seed 1234). derivative-free results follow their native query and stopping limits; the reporting point does not imply an equal query budget.}
\label{tab:iris_100_seed1234}
\centering
\setlength{\tabcolsep}{6pt}
\begin{tabular*}{\textwidth}{@{\extracolsep{\fill}}clccc@{}}
\toprule
\textbf{Rank} & \textbf{Optimizer} & \textbf{Best Loss} & \makecell{\textbf{Train}\\\textbf{Accuracy (\%)}} & \makecell{\textbf{Test}\\\textbf{Accuracy (\%)}} \\
\midrule
1 & FiniteDiff-Adam & $0.030639$ & $96.25$ & $100.00$ \\
2 & FiniteDiff-RMSprop & $0.024132$ & $96.25$ & $100.00$ \\
3 & PGPE-Adam & $0.043825$ & $95.00$ & $100.00$ \\
4 & PGPE-RMSprop & $0.041270$ & $100.00$ & $100.00$ \\
5 & PSR-Adam & $0.020250$ & $97.50$ & $100.00$ \\
6 & OCP-Adam & $0.023616$ & $97.50$ & $100.00$ \\
7 & PSR-RMSprop & $0.014761$ & $96.25$ & $100.00$ \\
8 & OCP-RMSprop & $0.017007$ & $98.75$ & $100.00$ \\
9 & Bayesian & $0.077353$ & $96.25$ & $95.00$ \\
10 & SPSA-Adam & $0.056108$ & $98.75$ & $95.00$ \\
11 & SPSA-RMSprop & $0.054352$ & $97.50$ & $95.00$ \\
12 & OCP-SGD & $0.051098$ & $90.00$ & $90.00$ \\
13 & COBYLA & $0.064264$ & $77.50$ & $80.00$ \\
14 & FiniteDiff-SGD & $0.078691$ & $65.00$ & $70.00$ \\
15 & SPSA-SGD & $0.064983$ & $70.00$ & $70.00$ \\
16 & PGPE-SGD & $0.064649$ & $71.25$ & $70.00$ \\
17 & PSR-SGD & $0.079693$ & $67.50$ & $70.00$ \\
18 & Nelder-Mead & $0.085969$ & $61.25$ & $60.00$ \\
\bottomrule
\end{tabular*}
\end{table}

\begin{table}[htbp]
\caption{Averaged performance metrics for the 4-qubit Iris classification task across three random seeds (123, 1234, 42) at the 100 step modular reporting point. Ordered by Average Test Accuracy. derivative-free results follow their native query and stopping limits; the reporting point does not imply an equal query budget.}
\label{tab:iris_average}
\centering
\setlength{\tabcolsep}{6pt}
\begin{tabular*}{\textwidth}{@{\extracolsep{\fill}}clccc@{}}
\toprule
\textbf{Rank} & \textbf{Optimizer} & \makecell{\textbf{Avg.}\\\textbf{Best Loss}} & \makecell{\textbf{Avg. Train}\\\textbf{Acc. (\%)}} & \makecell{\textbf{Avg. Test}\\\textbf{Acc. (\%)}} \\
\midrule
1 & OCP-RMSprop & $0.030933$ & $98.75$ & $100.00$ \\
2 & PSR-RMSprop & $0.030594$ & $97.92$ & $98.33$ \\
3 & FiniteDiff-RMSprop & $0.037270$ & $97.92$ & $98.33$ \\
4 & OCP-Adam & $0.038836$ & $98.33$ & $96.67$ \\
5 & PSR-Adam & $0.032819$ & $98.33$ & $95.00$ \\
6 & FiniteDiff-Adam & $0.044937$ & $97.08$ & $95.00$ \\
7 & SPSA-Adam & $0.073903$ & $97.08$ & $93.33$ \\
8 & SPSA-RMSprop & $0.057841$ & $97.50$ & $91.67$ \\
9 & PGPE-Adam & $0.049748$ & $93.75$ & $90.00$ \\
10 & OCP-SGD & $0.065094$ & $94.17$ & $88.33$ \\
11 & PGPE-RMSprop & $0.043929$ & $92.50$ & $86.67$ \\
12 & COBYLA & $0.089864$ & $75.83$ & $81.67$ \\
13 & Bayesian & $0.096988$ & $80.42$ & $76.67$ \\
14 & PSR-SGD & $0.123858$ & $67.08$ & $75.00$ \\
15 & SPSA-SGD & $0.130635$ & $62.08$ & $75.00$ \\
16 & PGPE-SGD & $0.103842$ & $70.42$ & $73.33$ \\
17 & FiniteDiff-SGD & $0.124271$ & $64.58$ & $73.33$ \\
18 & Nelder-Mead & $0.146813$ & $57.50$ & $61.67$ \\
\bottomrule
\end{tabular*}
\end{table}

\FloatBarrier

\subsection{$H_2$ VQE benchmark tables}\label{subsec:appendix_vqe}

\begin{table}[htbp]
\caption{Best observed energy and comparison with the 1.6~mHa threshold for the $H_2$ VQE task in 256-shot simulation (100 step modular reporting point and native derivative-free limits, seed 42). Best Energy is the minimum sampled energy of each reported trajectory. Ordering uses its absolute error from FCI, while the error column gives the signed difference.}
\label{tab:vqe_100}
\centering
\setlength{\tabcolsep}{4pt}
\begin{tabular*}{\textwidth}{@{\extracolsep{\fill}}clcccc@{}}
\toprule
\textbf{Rank} & \textbf{Optimizer} & \textbf{Best Energy (Ha)} & \makecell{\textbf{Signed}\\\textbf{Error (mHa)}} & \makecell{\textbf{Within}\\\textbf{1.6 mHa}} & \textbf{Steps/Queries} \\
\midrule
1 & PGPE-SGD & $-1.13728088$ & $-0.0107$ & Yes & $100$ \\
2 & PGPE-Adam & $-1.13734074$ & $-0.0706$ & Yes & $100$ \\
3 & SPSA-Adam & $-1.13743845$ & $-0.1683$ & Yes & $100$ \\
4 & SPSA-SGD & $-1.13760005$ & $-0.3299$ & Yes & $100$ \\
5 & FiniteDiff-RMSprop & $-1.13647501$ & $+0.7952$ & Yes & $100$ \\
6 & PSR-RMSprop & $-1.13812567$ & $-0.8555$ & Yes & $100$ \\
7 & PGPE-RMSprop & $-1.13814661$ & $-0.8764$ & Yes & $100$ \\
8 & OCP-RMSprop & $-1.13535290$ & $+1.9173$ & No & $100$ \\
9 & Bayesian & $-1.13493896$ & $+2.3312$ & No & $100$ \\
10 & PSR-SGD & $-1.13969957$ & $-2.4294$ & No & $100$ \\
11 & OCP-SGD & $-1.14005365$ & $-2.7835$ & No & $100$ \\
12 & FiniteDiff-SGD & $-1.14133333$ & $-4.0632$ & No & $100$ \\
13 & PSR-Adam & $-1.14133333$ & $-4.0632$ & No & $100$ \\
14 & FiniteDiff-Adam & $-1.13297608$ & $+4.2941$ & No & $100$ \\
15 & OCP-Adam & $-1.14186298$ & $-4.5928$ & No & $100$ \\
16 & SPSA-RMSprop & $-1.14202055$ & $-4.7504$ & No & $100$ \\
17 & COBYLA & $-1.13218418$ & $+5.0860$ & No & $32$ \\
18 & Nelder-Mead & $-1.14556135$ & $-8.2912$ & No & $93$ \\
\bottomrule
\end{tabular*}
\end{table}

\begin{sidewaystable}[p]
\caption{Resource utilization for the $H_2$ VQE task in 256-shot simulation at the 100 step modular reporting point and native derivative-free limits, seed 42. ``Steps/Queries'' denotes optimization steps for modular methods and objective queries for derivative-free methods. Logical-circuit and requested-shot totals are reported as N/R (not recorded) because the number of commuting measurement groups required for each Hamiltonian evaluation was not retained and therefore cannot be reconstructed reliably. The final column reports the first iteration at which the absolute change between two consecutive recorded objective values was below $10^{-4}$; it is a retrospective trajectory diagnostic and not a convergence criterion.}
\label{tab:vqe_resources}
\centering
\setlength{\tabcolsep}{3pt}
\begin{small}
\begin{tabular*}{\textheight}{@{\extracolsep{\fill}}lccccccc@{}}
\toprule
\textbf{Optimizer} & \makecell{\textbf{Final}\\\textbf{Energy (Ha)}} & \textbf{Steps/Queries} & \makecell{\textbf{Objective}\\\textbf{Evals}} & \makecell{\textbf{Logical}\\\textbf{Circuits}} & \makecell{\textbf{Requested}\\\textbf{Shots}} & \textbf{Time (s)} & \makecell{\textbf{First Small-Change}\\\textbf{Iteration}} \\
\midrule
COBYLA & $-1.132184$ & $32$ & $32$ & N/R & N/R & $5.78$ & $26$ \\
Nelder-Mead & $-1.145561$ & $93$ & $93$ & N/R & N/R & $19.19$ & $62$ \\
Bayesian & $-1.134939$ & $100$ & $100$ & N/R & N/R & $133.70$ & $1$ \\
FiniteDiff-SGD & $-1.128254$ & $100$ & $700$ & N/R & N/R & $145.02$ & $39$ \\
FiniteDiff-Adam & $-1.128608$ & $100$ & $700$ & N/R & N/R & $167.01$ & $3$ \\
FiniteDiff-RMSprop & $-1.136475$ & $100$ & $700$ & N/R & N/R & $128.44$ & -- \\
SPSA-SGD & $-1.131206$ & $100$ & $300$ & N/R & N/R & $63.17$ & $35$ \\
SPSA-Adam & $-1.132384$ & $100$ & $300$ & N/R & N/R & $66.34$ & $23$ \\
SPSA-RMSprop & $-1.126704$ & $100$ & $300$ & N/R & N/R & $64.95$ & $46$ \\
PGPE-SGD & $-1.129239$ & $100$ & $1{,}100$ & N/R & N/R & $206.98$ & $20$ \\
PGPE-Adam & $-1.132071$ & $100$ & $1{,}100$ & N/R & N/R & $197.41$ & $19$ \\
PGPE-RMSprop & $-1.129789$ & $100$ & $1{,}100$ & N/R & N/R & $192.70$ & $38$ \\
PSR-SGD & $-1.129239$ & $100$ & $2{,}500$ & N/R & N/R & $399.52$ & $15$ \\
OCP-SGD & $-1.127468$ & $100$ & $16{,}900$ & N/R & N/R & $3154.70$ & $24$ \\
PSR-Adam & $-1.129239$ & $100$ & $2{,}500$ & N/R & N/R & $439.79$ & $24$ \\
OCP-Adam & $-1.129239$ & $100$ & $16{,}900$ & N/R & N/R & $3575.86$ & $39$ \\
PSR-RMSprop & $-1.129456$ & $100$ & $2{,}500$ & N/R & N/R & $469.81$ & $37$ \\
OCP-RMSprop & $-1.129810$ & $100$ & $16{,}900$ & N/R & N/R & $4043.69$ & $38$ \\
\bottomrule
\end{tabular*}
\end{small}
\end{sidewaystable}

\begin{table}[htbp]
\caption{Best observed energy and error metrics for the 4-qubit $\text{H}_2$ VQE task in 256-shot simulation (100 step modular reporting point, Seed 123). Exact FCI ground state energy is $-1.137270$ Ha. derivative-free results follow their native query and stopping limits; the reporting point does not imply an equal query budget. Best Energy is the minimum sampled energy of each reported trajectory. Ordering uses its absolute error from FCI, while the error column gives the signed difference.}
\label{tab:vqe_100_seed123}
\centering
\setlength{\tabcolsep}{6pt}
\begin{tabular*}{\textwidth}{@{\extracolsep{\fill}}clcc@{}}
\toprule
\textbf{Rank} & \textbf{Optimizer} & \textbf{Best Energy (Ha)} & \textbf{Signed Error (mHa)} \\
\midrule
1 & OCP-SGD & $-1.14353860$ & $-6.2684$ \\
2 & FiniteDiff-RMSprop & $-1.14438743$ & $-7.1173$ \\
3 & FiniteDiff-SGD & $-1.14460084$ & $-7.3307$ \\
4 & FiniteDiff-Adam & $-1.14460084$ & $-7.3307$ \\
5 & PSR-SGD & $-1.14460084$ & $-7.3307$ \\
6 & SPSA-SGD & $-1.14729684$ & $-10.0267$ \\
7 & PGPE-RMSprop & $-1.14955796$ & $-12.2878$ \\
8 & SPSA-RMSprop & $-1.14991204$ & $-12.6419$ \\
9 & PSR-RMSprop & $-1.15026612$ & $-12.9959$ \\
10 & OCP-RMSprop & $-1.15062020$ & $-13.3500$ \\
11 & Bayesian & $-1.12366827$ & $+13.6019$ \\
12 & PGPE-SGD & $-1.15097428$ & $-13.7041$ \\
13 & COBYLA & $-1.15132836$ & $-14.0582$ \\
14 & PSR-Adam & $-1.15132836$ & $-14.0582$ \\
15 & SPSA-Adam & $-1.15168244$ & $-14.4123$ \\
16 & OCP-Adam & $-1.15203652$ & $-14.7663$ \\
17 & PGPE-Adam & $-1.15203652$ & $-14.7663$ \\
18 & Nelder-Mead & $-1.15239060$ & $-15.1204$ \\
\bottomrule
\end{tabular*}
\end{table}

\begin{table}[htbp]
\caption{Best observed energy and error metrics for the 4-qubit $\text{H}_2$ VQE task in 256-shot simulation (100 step modular reporting point, Seed 1234). Exact FCI ground state energy is $-1.137270$ Ha. derivative-free results follow their native query and stopping limits; the reporting point does not imply an equal query budget. Best Energy is the minimum sampled energy of each reported trajectory. Ordering uses its absolute error from FCI, while the error column gives the signed difference.}
\label{tab:vqe_100_seed1234}
\centering
\setlength{\tabcolsep}{6pt}
\begin{tabular*}{\textwidth}{@{\extracolsep{\fill}}clcc@{}}
\toprule
\textbf{Rank} & \textbf{Optimizer} & \textbf{Best Energy (Ha)} & \textbf{Signed Error (mHa)} \\
\midrule
1 & FiniteDiff-Adam & $-1.13744247$ & $-0.1723$ \\
2 & FiniteDiff-RMSprop & $-1.13744247$ & $-0.1723$ \\
3 & PGPE-RMSprop & $-1.13744247$ & $-0.1723$ \\
4 & Nelder-Mead & $-1.13815063$ & $-0.8805$ \\
5 & FiniteDiff-SGD & $-1.13618373$ & $+1.0864$ \\
6 & SPSA-RMSprop & $-1.13580872$ & $+1.4615$ \\
7 & COBYLA & $-1.13567208$ & $+1.5981$ \\
8 & SPSA-Adam & $-1.13474245$ & $+2.5277$ \\
9 & OCP-Adam & $-1.13405925$ & $+3.2109$ \\
10 & PSR-SGD & $-1.13403429$ & $+3.2359$ \\
11 & OCP-SGD & $-1.13368021$ & $+3.5900$ \\
12 & PGPE-SGD & $-1.13332613$ & $+3.9440$ \\
13 & PSR-Adam & $-1.13199464$ & $+5.2755$ \\
14 & SPSA-SGD & $-1.13177317$ & $+5.4970$ \\
15 & PGPE-Adam & $-1.13158069$ & $+5.6895$ \\
16 & OCP-RMSprop & $-1.13158069$ & $+5.6895$ \\
17 & PSR-RMSprop & $-1.13149989$ & $+5.7703$ \\
18 & Bayesian & $-1.12937542$ & $+7.8948$ \\
\bottomrule
\end{tabular*}
\end{table}

\begin{table}[htbp]
\caption{Averaged absolute error from exact FCI for the 4-qubit $\text{H}_2$ VQE task across three random seeds (123, 1234, 42) at the 100 step modular reporting point. derivative-free results follow their native query and stopping limits; the reporting point does not imply an equal query budget. Absolute errors are calculated from each seed's minimum sampled energy before averaging.}
\label{tab:vqe_average}
\centering
\setlength{\tabcolsep}{6pt}
\begin{tabular}{clc}
\toprule
\textbf{Rank} & \textbf{Optimizer} & \textbf{Avg. Absolute Error (mHa)} \\
\midrule
1 & FiniteDiff-RMSprop & $2.6949$ \\
2 & FiniteDiff-Adam & $3.9324$ \\
3 & FiniteDiff-SGD & $4.1601$ \\
4 & OCP-SGD & $4.2140$ \\
5 & PSR-SGD & $4.3320$ \\
6 & PGPE-RMSprop & $4.4455$ \\
7 & SPSA-SGD & $5.2845$ \\
8 & SPSA-Adam & $5.7028$ \\
9 & PGPE-SGD & $5.8863$ \\
10 & SPSA-RMSprop & $6.2846$ \\
11 & PSR-RMSprop & $6.5406$ \\
12 & PGPE-Adam & $6.8421$ \\
13 & COBYLA & $6.9141$ \\
14 & OCP-RMSprop & $6.9856$ \\
15 & OCP-Adam & $7.5233$ \\
16 & PSR-Adam & $7.7990$ \\
17 & Bayesian & $7.9426$ \\
18 & Nelder-Mead & $8.0974$ \\
\bottomrule
\end{tabular}
\end{table}

\FloatBarrier

\subsection{MNIST QCNN benchmark tables}\label{subsec:appendix_qcnn}

QCNN training uses mini batches of 16 samples for both modular and
derivative-free methods. Best and terminal losses refer to the recorded
mini batch objectives. Train and test accuracies are evaluated separately
at the selected parameters over the full training and test sets.

\begin{table}[htbp]
\caption{Best observed loss and accuracy metrics for the eight-qubit MNIST QCNN classification task in 256-shot simulation (100 step modular reporting point and native derivative-free limits, seed 42)}
\label{tab:qcnn_100_top}
\centering
\setlength{\tabcolsep}{6pt}
\begin{tabular*}{\textwidth}{@{\extracolsep{\fill}}clccc@{}}
\toprule
\textbf{Rank} & \textbf{Optimizer} & \textbf{Best Loss} & \makecell{\textbf{Train}\\\textbf{Accuracy (\%)}} & \makecell{\textbf{Test}\\\textbf{Accuracy (\%)}} \\
\midrule
1 & OCP-Adam & $0.082917$ & $94.38$ & $100.00$ \\
2 & OCP-RMSprop & $0.075435$ & $93.12$ & $100.00$ \\
3 & COBYLA & $0.103096$ & $93.75$ & $97.50$ \\
4 & PGPE-Adam & $0.134379$ & $92.50$ & $97.50$ \\
5 & PSR-Adam & $0.091747$ & $91.88$ & $97.50$ \\
6 & FiniteDiff-RMSprop & $0.138096$ & $91.88$ & $95.00$ \\
7 & OCP-SGD & $0.171259$ & $89.38$ & $95.00$ \\
8 & PSR-RMSprop & $0.070746$ & $88.12$ & $92.50$ \\
9 & FiniteDiff-Adam & $0.161526$ & $89.38$ & $90.00$ \\
10 & SPSA-Adam & $0.170758$ & $91.25$ & $90.00$ \\
11 & SPSA-RMSprop & $0.145649$ & $83.12$ & $90.00$ \\
12 & PGPE-RMSprop & $0.116093$ & $86.88$ & $90.00$ \\
13 & Nelder-Mead & $0.102530$ & $81.25$ & $80.00$ \\
14 & Bayesian & $0.194804$ & $70.62$ & $77.50$ \\
15 & FiniteDiff-SGD & $0.180711$ & $51.25$ & $47.50$ \\
16 & SPSA-SGD & $0.186819$ & $51.88$ & $47.50$ \\
17 & PGPE-SGD & $0.196306$ & $52.50$ & $47.50$ \\
18 & PSR-SGD & $0.178460$ & $51.25$ & $47.50$ \\
\bottomrule
\end{tabular*}
\end{table}

\begin{sidewaystable}[p]
\caption{Resource utilization for the eight-qubit MNIST QCNN classifier in 256-shot simulation at the 100 step modular reporting point and native derivative-free limits, seed 42. ``Steps/Queries'' denotes optimization steps for modular methods and objective queries for derivative-free methods. The final column reports the first iteration at which the absolute change between two consecutive recorded objective values was below $10^{-4}$; it is a retrospective trajectory diagnostic and not a convergence criterion.}
\label{tab:qcnn_resources}
\centering 
\setlength{\tabcolsep}{2pt} 
\begin{small} 
\begin{tabular*}{\textheight}{@{\extracolsep{\fill}}lccccccc@{}}
\toprule
\textbf{Optimizer} & \makecell{\textbf{Final}\\\textbf{Loss}} & \textbf{Steps/Queries} & \makecell{\textbf{Objective}\\\textbf{Evals}} & \makecell{\textbf{Logical}\\\textbf{Circuits}} & \makecell{\textbf{Requested}\\\textbf{Shots}} & \textbf{Time (s)} & \makecell{\textbf{First Small-Change}\\\textbf{Iteration}} \\
\midrule
COBYLA & $0.103096$ & $595$ & $595$ & $9{,}520$ & $2{,}437{,}120$ & $254.40$ & $49$ \\
Nelder-Mead & $0.107385$ & $1107$ & $1{,}107$ & $17{,}712$ & $4{,}534{,}272$ & $1057.45$ & $12$ \\
Bayesian & $0.194804$ & $1000$ & $1{,}000$ & $16{,}000$ & $4{,}096{,}000$ & $17304.97$ & $3$ \\
FiniteDiff-SGD & $0.220952$ & $100$ & $12{,}700$ & $203{,}200$ & $52{,}019{,}200$ & $2002.67$ & -- \\
FiniteDiff-Adam & $0.161526$ & $100$ & $12{,}700$ & $203{,}200$ & $52{,}019{,}200$ & $972.29$ & -- \\
FiniteDiff-RMSprop & $0.162896$ & $100$ & $12{,}700$ & $203{,}200$ & $52{,}019{,}200$ & $978.58$ & $97$ \\
SPSA-SGD & $0.237542$ & $100$ & $300$ & $4{,}800$ & $1{,}228{,}800$ & $28.02$ & -- \\
SPSA-Adam & $0.175162$ & $100$ & $300$ & $4{,}800$ & $1{,}228{,}800$ & $28.17$ & -- \\
SPSA-RMSprop & $0.148312$ & $100$ & $300$ & $4{,}800$ & $1{,}228{,}800$ & $27.65$ & -- \\
PGPE-SGD & $0.250035$ & $100$ & $1{,}100$ & $17{,}600$ & $4{,}505{,}600$ & $88.38$ & -- \\
PGPE-Adam & $0.170164$ & $100$ & $1{,}100$ & $17{,}600$ & $4{,}505{,}600$ & $88.40$ & -- \\
PGPE-RMSprop & $0.153953$ & $100$ & $1{,}100$ & $17{,}600$ & $4{,}505{,}600$ & $90.49$ & -- \\
PSR-SGD & $0.221915$ & $100$ & $12{,}700$ & $203{,}200$ & $52{,}019{,}200$ & $794.26$ & -- \\
OCP-SGD & $0.176391$ & $100$ & $25{,}300$ & $404{,}800$ & $103{,}628{,}800$ & $1722.06$ & -- \\
PSR-Adam & $0.107216$ & $100$ & $12{,}700$ & $203{,}200$ & $52{,}019{,}200$ & $795.80$ & -- \\
OCP-Adam & $0.095764$ & $100$ & $25{,}300$ & $404{,}800$ & $103{,}628{,}800$ & $1707.91$ & -- \\
PSR-RMSprop & $0.097957$ & $100$ & $12{,}700$ & $203{,}200$ & $52{,}019{,}200$ & $827.06$ & -- \\
OCP-RMSprop & $0.085378$ & $100$ & $25{,}300$ & $404{,}800$ & $103{,}628{,}800$ & $1720.14$ & -- \\
\bottomrule
\end{tabular*}
\end{small}
\end{sidewaystable}

\begin{table}[htbp]
\caption{Best observed loss and accuracy metrics for the 8-qubit MNIST QCNN task in 256-shot simulation (100 step modular reporting point, Seed 123). derivative-free results follow their native query and stopping limits; the reporting point does not imply an equal query budget.}
\label{tab:qcnn_100_seed123}
\centering
\setlength{\tabcolsep}{6pt}
\begin{tabular*}{\textwidth}{@{\extracolsep{\fill}}clccc@{}}
\toprule
\textbf{Rank} & \textbf{Optimizer} & \textbf{Best Loss} & \makecell{\textbf{Train}\\\textbf{Accuracy (\%)}} & \makecell{\textbf{Test}\\\textbf{Accuracy (\%)}} \\
\midrule
1 & PSR-RMSprop & $0.052925$ & $94.38$ & $97.50$ \\
2 & OCP-RMSprop & $0.054062$ & $93.75$ & $97.50$ \\
3 & FiniteDiff-Adam & $0.164983$ & $92.50$ & $95.00$ \\
4 & OCP-Adam & $0.053757$ & $92.50$ & $95.00$ \\
5 & FiniteDiff-RMSprop & $0.132205$ & $85.00$ & $92.50$ \\
6 & SPSA-RMSprop & $0.144800$ & $93.12$ & $92.50$ \\
7 & PGPE-Adam & $0.121453$ & $91.88$ & $92.50$ \\
8 & PSR-Adam & $0.049915$ & $91.88$ & $92.50$ \\
9 & SPSA-Adam & $0.156280$ & $90.62$ & $90.00$ \\
10 & PGPE-RMSprop & $0.108904$ & $91.25$ & $90.00$ \\
11 & OCP-SGD & $0.148003$ & $81.25$ & $82.50$ \\
12 & FiniteDiff-SGD & $0.169921$ & $69.38$ & $65.00$ \\
13 & PGPE-SGD & $0.182979$ & $68.75$ & $65.00$ \\
14 & PSR-SGD & $0.173003$ & $69.38$ & $65.00$ \\
15 & SPSA-SGD & $0.176621$ & $70.00$ & $62.50$ \\
16 & Nelder-Mead & $0.181351$ & $65.00$ & $60.00$ \\
17 & COBYLA & $0.136913$ & $53.12$ & $42.50$ \\
18 & Bayesian & $0.184438$ & $55.62$ & $42.50$ \\
\bottomrule
\end{tabular*}
\end{table}

\begin{table}[htbp]
\caption{Best observed loss and accuracy metrics for the 8-qubit MNIST QCNN task in 256-shot simulation (100 step modular reporting point, Seed 1234). derivative-free results follow their native query and stopping limits; the reporting point does not imply an equal query budget.}
\label{tab:qcnn_100_seed1234}
\centering
\setlength{\tabcolsep}{6pt}
\begin{tabular*}{\textwidth}{@{\extracolsep{\fill}}clccc@{}}
\toprule
\textbf{Rank} & \textbf{Optimizer} & \textbf{Best Loss} & \makecell{\textbf{Train}\\\textbf{Accuracy (\%)}} & \makecell{\textbf{Test}\\\textbf{Accuracy (\%)}} \\
\midrule
1 & OCP-RMSprop & $0.074230$ & $95.00$ & $92.50$ \\
2 & FiniteDiff-Adam & $0.151816$ & $93.12$ & $90.00$ \\
3 & PSR-Adam & $0.072979$ & $93.75$ & $90.00$ \\
4 & OCP-Adam & $0.067068$ & $96.25$ & $90.00$ \\
5 & COBYLA & $0.168972$ & $91.25$ & $87.50$ \\
6 & FiniteDiff-RMSprop & $0.152122$ & $95.62$ & $87.50$ \\
7 & PGPE-Adam & $0.113118$ & $95.00$ & $87.50$ \\
8 & PGPE-RMSprop & $0.110376$ & $96.25$ & $87.50$ \\
9 & PSR-RMSprop & $0.075025$ & $96.25$ & $87.50$ \\
10 & SPSA-RMSprop & $0.169268$ & $89.38$ & $82.50$ \\
11 & OCP-SGD & $0.173596$ & $91.25$ & $82.50$ \\
12 & SPSA-Adam & $0.150545$ & $92.50$ & $80.00$ \\
13 & FiniteDiff-SGD & $0.212608$ & $69.38$ & $75.00$ \\
14 & Bayesian & $0.213774$ & $78.12$ & $70.00$ \\
15 & SPSA-SGD & $0.210176$ & $74.38$ & $70.00$ \\
16 & PSR-SGD & $0.213262$ & $71.88$ & $70.00$ \\
17 & PGPE-SGD & $0.212899$ & $64.38$ & $60.00$ \\
18 & Nelder-Mead & $0.235217$ & $61.88$ & $57.50$ \\
\bottomrule
\end{tabular*}
\end{table}

\begin{table}[htbp]
\caption{Averaged performance metrics for the 8-qubit MNIST QCNN classification task across three random seeds (123, 1234, 42) at the 100 step modular reporting point. Ordered by Average Test Accuracy. derivative-free results follow their native query and stopping limits; the reporting point does not imply an equal query budget.}
\label{tab:qcnn_average}
\centering
\setlength{\tabcolsep}{6pt}
\begin{tabular*}{\textwidth}{@{\extracolsep{\fill}}clccc@{}}
\toprule
\textbf{Rank} & \textbf{Optimizer} & \makecell{\textbf{Avg.}\\\textbf{Best Loss}} & \makecell{\textbf{Avg. Train}\\\textbf{Acc. (\%)}} & \makecell{\textbf{Avg. Test}\\\textbf{Acc. (\%)}} \\
\midrule
1 & OCP-RMSprop & $0.067909$ & $93.96$ & $96.67$ \\
2 & OCP-Adam & $0.067914$ & $94.38$ & $95.00$ \\
3 & PSR-Adam & $0.071547$ & $92.50$ & $93.33$ \\
4 & PSR-RMSprop & $0.066232$ & $92.92$ & $92.50$ \\
5 & PGPE-Adam & $0.122983$ & $93.13$ & $92.50$ \\
6 & FiniteDiff-RMSprop & $0.140808$ & $90.83$ & $91.67$ \\
7 & FiniteDiff-Adam & $0.159442$ & $91.67$ & $91.67$ \\
8 & PGPE-RMSprop & $0.111791$ & $91.46$ & $89.17$ \\
9 & SPSA-RMSprop & $0.153239$ & $88.54$ & $88.33$ \\
10 & OCP-SGD & $0.164286$ & $87.29$ & $86.67$ \\
11 & SPSA-Adam & $0.159194$ & $91.46$ & $86.67$ \\
12 & COBYLA & $0.136327$ & $79.37$ & $75.83$ \\
13 & Nelder-Mead & $0.173033$ & $69.38$ & $65.83$ \\
14 & Bayesian & $0.197672$ & $68.12$ & $63.33$ \\
15 & FiniteDiff-SGD & $0.187747$ & $63.34$ & $62.50$ \\
16 & PSR-SGD & $0.188242$ & $64.17$ & $60.83$ \\
17 & SPSA-SGD & $0.191205$ & $65.42$ & $60.00$ \\
18 & PGPE-SGD & $0.197395$ & $61.88$ & $57.50$ \\
\bottomrule
\end{tabular*}
\end{table}

\FloatBarrier

Table~\ref{tab:best_overall_average_optimizers} summarizes the leading
mean configuration for each workload.

\begin{table}[htbp]
\caption{Configuration with the highest reported average metric across all three seeds (123, 1234, 42) for each quantum task at the 100 step modular reporting point. Classification tasks are ordered by Average Test Accuracy. derivative-free results follow their native query and stopping limits; the reporting point does not imply an equal query budget.}
\label{tab:best_overall_average_optimizers}
\centering
\setlength{\tabcolsep}{6pt}
\begin{tabular}{@{\extracolsep{\fill}}llrl@{}}
\toprule
\textbf{Task} & \textbf{Best Optimizer (Avg.)} & \multicolumn{2}{c}{\textbf{Average Performance Metric}} \\
\midrule
MaxCut QAOA & PGPE-SGD & $0.6549$ & (Distance to Target) \\
Iris Classification (4-qubit) & OCP-RMSprop & $100.00\%$ & (Average Test Acc.) \\
$\text{H}_2$ VQE (4-qubit) & FiniteDiff-RMSprop & $2.6949$ mHa & (Abs. Error from FCI) \\
MNIST QCNN (8-qubit) & OCP-RMSprop & $96.67\%$ & (Average Test Acc.) \\
\bottomrule
\end{tabular}
\end{table}

\section{QCNN Architecture and Gate Templates}\label{secE1}
The input is encoded by Hadamard gates followed by $P(2x_j)$ on each of the eight qubits, as in Eq.~\eqref{eq:mnist_ref}. The circuit uses convolution and pooling blocks with three independent parameters per block. Parameters are not shared between blocks.

\FloatBarrier

\subsection{Convolution and pooling blocks}
Let $C_{i\to j}$ denote a CNOT with control $i$ and target $j$. For an ordered pair $(i,j)$, the convolution gates are applied in the following order:
\[
R_Z^{(j)}(-\pi/2),\ C_{j\to i},\ R_Z^{(i)}(a),\ R_Y^{(j)}(b),\ C_{i\to j},\ R_Y^{(j)}(c),\ C_{j\to i},\ R_Z^{(i)}(\pi/2).
\]
The pooling block applies the first six gates of this sequence. Its source wire receives no subsequent gates. Tracing out the inactive wires describes the reduced state of the surviving register; that state need not be pure.

\subsection{Wire schedule and parameter count}
The first stage applies convolutions on $(0,1),(2,3),(4,5),(6,7),(1,2),(3,4),(5,6),(7,0)$ and pooling on $(0,4),(1,5),(2,6),(3,7)$. The surviving wires are $(4,5,6,7)$. The second stage applies convolutions on $(4,5),(6,7),(5,6),(7,4)$ and pooling on $(4,6),(5,7)$. The last stage applies convolutions on $(6,7),(7,6)$ and pooling on $(6,7)$. The output observable is $Z_7$. The 14 convolution blocks and seven pooling blocks contain $14\times3+7\times3=63$ parameters and $14\times3+7\times2=56$ CNOTs before transpilation.

\section{Observable Curvature Used by OCP}\label{secF1}

The shift rule for second derivatives is obtained by applying the first
derivative rule twice \citep{Mari2021Estimating}. Let $f(\boldsymbol\phi)$ be the
objective expressed in independent elementary gate angles, and let
$\mathbf e_\ell$ now denote a basis vector in that gate angle space. For
$s=\pi/2$, define
\begin{align}
 \mathcal S_\ell f(\boldsymbol\phi)
 &=\frac{f(\boldsymbol\phi+s\mathbf e_\ell)-
 f(\boldsymbol\phi-s\mathbf e_\ell)}2,\\
 \mathcal S_\ell\mathcal S_m f(\boldsymbol\phi)
 &=\frac14\sum_{u,v\in\{-1,+1\}}uv\,
 f(\boldsymbol\phi+us\mathbf e_\ell+vs\mathbf e_m).
 \label{eq:ocp_gate_second_shift}
\end{align}
For affine gate angles, the energy implementation assembles
\begin{equation}
 \partial_j^2J=\sum_{\ell,m}a_{\ell j}a_{mj}
 \mathcal S_\ell\mathcal S_m f.
 \label{eq:ocp_assembled_diagonal}
\end{equation}
The terms with $\ell=m$ use offsets $0$ and $\pm\pi$ on one gate;
terms with $\ell\ne m$ shift two gates. This explains why the energy cost depends
on gate occurrences rather than solely on the number of trainable parameters.
The implemented measurements replace each exact expectation by an estimate.

For a parameter that appears in one Pauli rotation, the exact observable expectation satisfies
\begin{equation}
h_{bj}=\partial_j^2z_b=\frac{z_b(\boldsymbol\theta+\pi\mathbf e_j)+z_b(\boldsymbol\theta-\pi\mathbf e_j)-2z_b(\boldsymbol\theta)}4.
\end{equation}
The classifier denominator is the batch mean of $|h_{bj}|$ plus $\epsilon_{\mathrm{reg}}$. The implemented estimates use finite measurements. By contrast, the MSE Hessian diagonal is
\begin{equation}
\partial_j^2\mathcal L=\frac1B\sum_b\left[\frac12(\partial_jz_b)^2+(p_b-y_b)\partial_j^2z_b\right].
\end{equation}
The Gauss--Newton diagonal retains only the squared derivative term. Neither expression equals the OCP denominator. The denominator rescales the direction; it does not change or smooth the objective landscape. For energy tasks, OCP uses the absolute objective Hessian diagonal plus regularization, including mixed gate derivatives through the chain rule.

In \texttt{OCPGradient}, the first derivative shift plan is applied twice for
each original parameter. This includes all pairs of elementary gates controlled
by that parameter, and hence the mixed gate derivatives in
Eq.~\eqref{eq:diagonal_qfim_main}. It does not construct the full Hessian with
respect to distinct original parameters. Contributions at identical gate offset
vectors are combined, and each distinct vector is evaluated once per iteration.
The zero offset supplies the monitored objective. The resulting diagonal and
PSR gradient are combined componentwise; no matrix inversion is performed.

In \texttt{OCPClassifierGradient}, each sample is evaluated at the unshifted
parameters and at four shifts per coordinate: $+\pi/2$, $-\pi/2$, $+\pi$, and
$-\pi$. The first pair supplies the PSR gradient, while the second pair and the
unshifted expectation supply the observable curvature. Absolute values are
taken separately for each sample before the batch average is formed. This gives
$B(1+4M)$ sample expectations, or 1040 for Iris and 4048 for QCNN per iteration.
The implementation retains both $\pi$ shifts as separate evaluations and does
not reduce this count using their equivalence for an ideal Pauli rotation.

\section{Hardware Execution Records}\label{sec:hardware_logs}
Table~\ref{tab:hardware_run_ids} identifies the session and training count for
each plotted hardware run. The MaxCut and $H_2$ sessions contain 26 jobs each
(one initial reference and 25 training batches), Iris contains 52 jobs
(50 training queries and separate training/test evaluations), and QCNN contains
11 jobs (ten training batches and one combined evaluation job). Including
reference and post-training evaluations, the archives contain 76, 900, 40,680,
and 276 returned expectation values for MaxCut, Iris, QCNN, and $H_2$,
respectively. For $H_2$, these are energy evaluations combining multiple Pauli
terms, not verified physical-circuit counts. Runtime batching and service-side
mitigation are distinct from the logical training counts.
\begin{table}[htbp]
\caption{Session identifiers for the reported runs on \texttt{ibm\_aachen}. Evaluation counts cover the complete training run and exclude initial reference and final accuracy evaluations.}
\label{tab:hardware_run_ids}
\centering
\small
\setlength{\tabcolsep}{4pt}
\renewcommand{\arraystretch}{1.15}
\begin{tabular}{@{}lll@{}}
\toprule
\makecell[l]{\textbf{Task and}\\\textbf{method}} & \makecell[l]{\textbf{Training}\\\textbf{evaluations}} & \textbf{Session identifier} \\
\midrule
\makecell[l]{MaxCut\\SPSA-SGD} & 75 objectives & \nolinkurl{a3c5c842-b509-4930-9efc-aa9aac4943aa} \\
\addlinespace
\makecell[l]{Iris\\COBYLA} & 800 sample circuits & \nolinkurl{1f9c0e26-da2c-45ae-8211-5c2b99cd87e9} \\
\addlinespace
\makecell[l]{QCNN\\OCP-RMSprop} & 40,480 sample circuits & \nolinkurl{14c4321e-9670-4571-a51d-a79f80a85f08} \\
\addlinespace
\makecell[l]{$H_2$ VQE\\PGPE-SGD} & 275 objectives & \nolinkurl{2e502439-33b1-460e-a070-f2122803e9fd} \\
\bottomrule
\end{tabular}
\end{table}

\end{appendices}
\end{document}